\documentclass[12pt,a4paper]{article}
\newcommand{\bra}[1]{\langle #1|}
\newcommand{\ket}[1]{|#1\rangle}
\newcommand{\braket}[2]{\langle #1|#2\rangle}
\begin{document}
\title{Teleportation and Superdense Coding with Genuine Quadripartite 
Entangled States}
\author{B. Pradhan, Pankaj Agrawal and A. K. Pati\footnote{email:
bpradhan@iopb.res.in,agrawal@iopb.res.in,akpati@iopb.res.in}\\
Institute of Physics \\ Sachivalaya Marg, Bhubaneswar, Orissa, India 751 005}
\maketitle

\begin{abstract}
We investigate the usefulness of different classes of genuine quadripartite
entangled states as quantum resources for teleportation and superdense 
coding. We examine the possibility of teleporting unknown one, two and three 
qubit states. We show that one can use the teleportation protocol
to send any general one and two qubit states. A restricted class of three
qubit states can also be faithfully teleported. We also explore
superdense coding protocol in single-receiver and multi-receiver scenarios.  
We show that there exist genuine quadripartite entangled states that can
be used to transmit four cbits by sending two qubits. We also discuss
some interesting features of multi-receiver scenario under LOCC paradigm.

\end{abstract}

\vspace{1in}

%

\newpage

\section{Introduction}
 
One of the intriguing feature of quantum mechanics is the quantum
entanglement. This feature has been exploited to do several amazing tasks 
which are otherwise impossible.  In particular, the entanglement can be 
used as a quantum resource to carry out a number of computational and 
information processing tasks. Such tasks include teleportation of an unknown 
quantum state \cite{bbcjpw}, superdense coding \cite{bw}, entanglement 
swapping \cite{zzhe}, remote state preparation \cite{akp,bdsstw}, 
secret sharing \cite{hbb}, quantum cryptography \cite{grtz} and many others.
Analysis of such quantum phenomena may allow us a better understanding
of the structure of the quantum mechanics framework.

  A quantum system may consist of two or more subsystems which 
may correspondingly have bipartite or multipartite
entanglement. Characterization and uses of bipartite entanglement 
are better understood than those of multipartite entanglement.
A number of protocols which were first introduced in the context of 
a bipartite system can be extended to a multipartite system. However,
in the case of multipartite systems, the entanglement environment is
quite complex and its nature is still not fully
understood. Such entangled states can be classified according to
different schemes. These classes exhibit different types of entanglement
properties. All classes may not be suitable for some of the information processing
tasks. The tripartite states have been classified according to 
stochastic local operation and classical communication (SLOCC)
into six categories. Two of these categories have genuine tripartite
entanglement, viz. GHZ-states and W-states \cite{dvc}. The utilities of
these states have been explored in a number of papers 
\cite{gtrz} $-$ \cite{yy}. 
The quadripartite states have also been classified 
according to SLOCC \cite{vddv}. There are nine categories. Some of
them have genuine quadripartite entanglement. But usefulness 
of the multipartite states beyond tripartite states is still to 
be explored in some details. Such studies may even allow a better 
understanding of multiparticle entanglement and classification of
quantum states according to their entanglement properties. 
It is worth mentioning that, a task-based classification scheme have been proposed by
 Bru\ss~ et al \cite{bdlmss, blssdm}  to classify mixed states and multipartite 
states according to their densecodabilty.

In this paper, we study various protocols for quantum teleportation and
superdense coding in the context of quadripartite entangled states. In this
scenario, there can be two, three, or four parties (Alice, Bob, Charlie
and Dennis). These parties share four particles in an entangled state.
We shall take these states to be genuinely quadripartite entangled
states in the sense that these states cannot be written as a direct
product of bipartite entangled states, or as a direct product of
a tripartite and a single-particle state.

  In the next section, we enumerate the quadripartite states of qubits
that we shall be considering. We shall explore the possibility of 
teleporting an unknown one-qubit, two-qubit, or three-qubit state. 
It is known (and we review it) that it is possible to teleport an unknown
one-qubit state using a variety of quadripartite states. One can do
this using a number of different protocols. These protocols can involve
only two-particle, or three-particle or four-particle von Neumann 
measurements. The situation is a bit more complicated in the case
of the teleportation of a general two-qubit state. Although some special
two-qubit states can be teleported by a number of different quadripartite
states, but a general state often cannot be. We show that a specific state
can be used to teleport a general two-qubit state. This state is different
from the one that was discussed in the literature \cite{yc}. A limited set of 
three-qubit states can also be teleported by using some of the 
quadripartite states which we discuss in the next section. However, to
teleport an arbitrary three-qubit state, one may need an appropriate
six-qubit entangled state.

Apart from the teleportation protocol, we discuss superdense coding using
quadripartite states as a quantum resource. We discuss two scenarios:
single-receiver and multi-receiver. In both the cases, there is just one sender. 
In the case of single-receiver scenario, there exist several possibilities: 
i) transmit two-cbits by sending one qubit, ii) transmit three, or 
four-cbits by sending two qubits, iii) transmit four-cbits by sending 
three qubits. Here the case of sending four cbits by sending two qubits 
is clearly more interesting. It turns out that there exist quadripartite
states that can be used as a quantum resource to accomplish the task of
transmitting four cbits by sending two qubits. More than four cbits cannot
be sent using quadripartite entangled state because the dimensionality
of the Hilbert space of four qubits is sixteen. We also discuss multi-receiver
scenarios in the framework of LOCC distinguishability of a set of orthogonal
states.


The plan of the paper is as follows. In section II, we enumerate the quadripartite
entangled states that we consider in this paper. In section III, we discuss the
use of these states as a quantum resource for the teleportation. In section IV, we discuss
the protocol of superdense coding. Finally, in section V, we present our conclusions.

\section{Quadripartite Entangled States}

The bipartite entangled states are the simplest of the entangled states. Such
states can be classified and their entanglement quantified. All bipartite
entangled states belong to one equivalence class under 
SLOCC. The representative state of this class can
be taken to be any of the Bell states. These Bell states are maximally entangled
states and can be used fruitfully for the teleportation and the superdense
coding as shown in the original papers that introduced these protocols \cite{bbcjpw,bw}.

One may think that if we go beyond bipartite entangled states to multipartite
entangled states, one may be able to accomplish tasks which are not otherwise
feasible. However, the nature of entanglement in the case of multipartite
entangled state is multifaceted and far from being understood. The genuine tripartite
entangled states have also been classified on the basis of SLOCC \cite{dvc}. There are 
two classes: i) the class with the representative state, $|GHZ \rangle$,
ii) the class with the representative state, $|W \rangle$. States belonging
to these classes can be used to successfully carry out the protocol of
teleportation and superdense coding \cite{gtrz,jpok,ap}.

Beyond tripartite entangled states, one may consider quadripartite entangled
states. On the basis of SLOCC, such states have also been classified. We consider
a set of states from this classification and consider the possibility of 
implementing the protocol of teleportation and superdense coding. These states are 
given by

\begin{eqnarray}
 \ket{Q1} &  \equiv  & \ket{GHZ}=\frac{1}{2}(\ket{0000}+\ket{1111}) \\
\ket{Q2}  & \equiv  & \ket{W}=\frac{1}{2}(\ket{0001}+\ket{0010}+\ket{0100}+\ket{1000}) \\
\ket{Q3}  & \equiv  &  \ket{\Omega}=\frac{1}{\sqrt{2}}(\ket{0}\ket{\varphi^+}\ket{0}+\ket{1}
\ket{\varphi^-}\ket{1})  \\
 \ket{Q4} &  = & \frac{1}{2}(\ket{0000}+\ket{0101}+\ket{1000}+\ket{1110})  \\
 \ket{Q5} &  = & \frac{1}{2}(\ket{0000}+\ket{1011}+\ket{1101}+\ket{1110})  
\end{eqnarray}

Here $\ket{\varphi^\pm}$ are Bell states defined in (15).
According to the classification of Verstrate et al \cite{vddv} quadripartite states
 $\ket{GHZ}$ and $\ket{\Omega}$ 
belong to $G_{abcd}$ class, the states $\ket{W}, \ket{Q4}$
 and $\ket{Q5}$ belong to $L_{ab_3}, L_{0_{5\oplus\overline{3}}}$ and
 $L_{0_{7\oplus\overline{1}}}$ respectively.  Among these
five entangled states $\ket{GHZ}$ and $\ket{W}$ states are
symmetric with respect to permutation of particles; thus any quantum
information task performed using these states are independent of distribution
of particles among the parties. For later three states it varies depending on
the distribution of particles among the parties. 

There are a number of ways to see that the above states have genuine
quadripartite entanglement. One way is to find out the states of the
system after tracing out one, two or three particles. If the state is mixed
in each case, then this would be an indication of genuine quadripartite
entanglement. This is what we observe. In addition we note the following.
The $\ket{GHZ}$ has a mixed 3-tangle \cite{ckw} of zero when one of the party is 
traced out. The mixed 3-tangle for $\ket{W}$ is zero and concurrence of 
1/2 when two qubits are traced out. For $\ket{Q4}$ mixed 3-tangle of 1/2 is obtained
if qubit 2,3 and 4 are traced out, zero when qubit 1 is traced out. By tracing
out qubit 1 and (3 or 4) one can get a concurrence equal to 1/2, while the other
concurrence vanish. The state $\ket{Q5}$ has a concurrences equal to zero if
two qubits are traced out and mixed 3-tangle of 1/2 if particle 2,3 and 4 are
traced out.  
The $\ket{\Omega}$  state is the cluster state introduced by Briegel and  
 Raussendorf  \cite{br, rb} . 
This state is considered to have maximum connectedness and high persistence of
entanglement and has been discussed extensively in the context of one-way
quantum computation. The concurrence of this state is zero with any of the two
qubits traced out.

\section{Teleportation}

    In this section, we consider the teleportation of the unknown states of one, two,
and three qubits. In the case of the teleportation the arbitrary state of one qubit,
a number of situations may exist. There may be just two parties, or more. There
is a possibility of making four-qubit, three-qubit, two-qubit, and one-qubit
von Neumann measurements or a combinations of them. As making
a measurement involving a larger number of qubits may be more
difficult, it would be interesting to know if the protocol would work
with the measurement on fewer particles. In the case of
the transmission of unknown two-qubit states, there can be 
situations of two or three parties with quadripartite entangled states, 
or Alice could have option of making different types of measurements.
In the case of transmitting an unknown three-qubit state, with
quadripartite states as a quantum resource, there can be only
two parties: Alice and Bob.

We look at some of the above
situations below. Depending on the number of parties, the classical
communication cost of the protocol would be different. If there are more
than two parties, then the number of transmitted cbits would increase
as the information about the measurement results would be distributed.
One issue in classical communication would be the number of cbits that
must be transmitted to Bob who wishes to convert the state of his qubit to the 
state of the unknown state of the qubit that Alice has. When there are 
only two parties, Alice and Bob, then Alice could encode in the cbits either
the results of her measurements or the unitary operations that
Bob should apply to his qubit. When Alice makes a series of measurements,
then the latter option is simpler. Of course, Alice and Bob would need
to have a prior understanding of the option that Alice would use. 

\subsection{Teleportation of a single-qubit state}

In  this scenario, Alice wishes to teleport an unknown qubit state
 $\ket{\psi}_a= \alpha \ket{0}+\beta \ket{1}$ to Bob. They share
a quantum channel given by one of the quadripartite states of
the last section. Of the four entangled qubits, Alice has qubits 1, 2, and 3 
and Bob has the qubit 4. As in the conventional teleportation protocol,
 Alice's strategy is to make von Neumann measurements involving particles
a, 1, 2, and 3 and communicate the results to Bob. Bob then performs necessary 
unitary operation on his qubit according to the received message to convert
the state of his qubit to that of the unknown qubit. As noted above, Alice has several
 choices of bases to perform the measurement. She may
choose a basis of four particles or successive two-particle Bell basis or
three-particle and one-particle basis for measurement. Let us now
discuss various states of the last section as a quantum resource.

\subsubsection{\bf Teleportation using $\ket{GHZ}$ state}

     Alice has the qubits a, 1, 2, and 3. Bob has the qubit 4. Alice wishes
   to teleport the unknown state of the particle a,
\begin{equation}
  \ket{\psi}_a= \alpha \ket{0}_a+\beta \ket{1}_a .
\end{equation}
Here $\alpha$ and $\beta$ are complex numbers.
 The combined state of the five-qubit systems can be written as:
\begin{equation}
\ket{\psi}_a \ket{GHZ}_{1234}=\frac{1}{\sqrt{2}} (\alpha \ket{0}_a+
 \beta \ket{1}_a)\otimes(\ket{0000}_{1234}+ \ket{1111}_{1234}).
\end{equation}

We can rewrite this combined state depending on the type of the measurement
Alice wishes to make. If Alice wishes to make four-particle von Neumann
measurement, then we can rewrite the above state as,

\begin{eqnarray}
\ket{\psi}_a \ket{GHZ}_{1234} & = & \frac{1}{\sqrt{2}}(\alpha \ket{0000}_{a123} \ket{0}_{4}+ \alpha \ket{0111}_{a123}\ket{1}_{4}+ \beta \ket{1000}_{a123}\ket{0}_{4}+
 \beta \ket{1111}_{a123}\ket{1}_{4} )                                                   \nonumber                  \\ 
& = & \frac{1}{2}[\ket{4GHZ_{1}^{+}}_{a123} (\alpha \ket{0}_{4}+ \beta \ket{1}_{4})+
           \ket{4GHZ_{1}^{-}}_{a123}(\alpha  \ket{0}_{4}-\beta \ket{1})_{4} + \nonumber  \\
   & &       \ket{4GHZ_{2}^{+}} _{a123}(\alpha \ket{1}_{4}+\beta  \ket{0})_{4}+
          \ket{4GHZ_{2}^{-}}_{a123}(\alpha \ket{1}_{4}-\beta \ket{0}_{4})],
\end{eqnarray}

where,
\begin{eqnarray}
\ket{4GHZ_{1}^{\pm}} & = & \frac{1}{\sqrt{2}}(\ket{0000}\pm\ket{1111}) \\
\ket{4GHZ_{2}^{\pm}} & = & \frac{1}{\sqrt{2}}(\ket{0111}\pm\ket{1000}).
\end{eqnarray}

According to the results of the measurement,  Alice sends two bits of  classical 
information to Bob,  encoding either the results of her measurements, or
the unitary operation that Bob should apply.   Bob performs one of the 
$\{\sigma_0, \sigma_1, i \sigma_2,\sigma_3\}$ operations
to convert the state of his qubit to that of the unknown qubit a. 

Instead of making a four-particle  von Neumann measurement,  Alice may 
wish to make a  three-particle  followed by one-particle von Neumann 
measurements. Or, there can be three parties, Alice, Bob, and Charlie. 
In this latter case, Alice may have the qubits a, 1, and 2, while Charlie
has the qubit 3 and Bob has the qubit 4. To see how the protocol
would work in this situation, we rewrite the combined state (8) as,

\begin{eqnarray}
\ket{\psi}_a \ket{GHZ}_{1234} & = & \frac{1}{\sqrt{2}}(\alpha \ket{000}_{a12} \ket{0}_{3} \ket{0}_{4}+ 
\alpha \ket{011}_{a12}  \ket{1}_{3} \ket{1}_{4}+ \beta \ket{100}_{a12}  \ket{0}_{3} \ket{0}_{4}+ \nonumber \\
 & & \beta \ket{111}_{a12}  \ket{1}_{3} \ket{1}_{4} )                            \nonumber  \\ 
& = & \frac{1}{2 \sqrt{2}}(\ket{3GHZ_{1}^{+}}_{a12} \ket{+}_{3} + \ket{3GHZ_{1}^{-}}_{a12} \ket{-}_{3})
                 (\alpha \ket{0}_{4}+ \beta \ket{1}_{4}) +  \nonumber \\
   & &    (\ket{3GHZ_{1}^{+}}_{a12} \ket{-}_{3} + \ket{3GHZ_{1}^{-}}_{a12} \ket{+}_{3})  
                (\alpha  \ket{0}_{4}-\beta \ket{1})_{4} + \nonumber  \\
   & &   (\ket{3GHZ_{2}^{+}}_{a12} \ket{+}_{3} - \ket{3GHZ_{2}^{-}}_{a12} \ket{-}_{3})  
                (\alpha \ket{1}_{4}+\beta  \ket{0})_{4}+ \nonumber \\
   & &    (\ket{-3GHZ_{2}^{+}}_{a12} \ket{-}_{3} + \ket{3GHZ_{2}^{-}}_{a12} \ket{+}_{3})    
                (\alpha \ket{1}_{4}-\beta \ket{0}_{4}),
\end{eqnarray}

where,
\begin{eqnarray}
\ket{3GHZ_{1}^{\pm}} & = & \frac{1}{\sqrt{2}}(\ket{000}\pm\ket{111}) \\
\ket{3GHZ_{2}^{\pm}} & = & \frac{1}{\sqrt{2}}(\ket{011}\pm\ket{100}).
\end{eqnarray}

and,

\begin{equation}
\ket{\pm}=\frac{1}{\sqrt{2}}(\ket{0}\pm\ket{1}).
\end{equation}

Here in the case of two parties, the simplest thing for the Alice would
be to encode in two cbits the unitary operation that Bob should apply 
on his qubit after she makes the measurement. So here classical 
communication cost would still be two cbits. However, in the scenario
of three parties, there will be need of three cbits of classical communication.
This communication could take many forms. Some examples are:
Alice sends two cbits and Charlie one cbit to Bob about the results of measurements;
Charlie sends one cbit to Alice, who then sends two cbits to Bob, encoding
the unitary operation that Bob should apply. In all cases Bob will make
one of the $\{\sigma_0, \sigma_1, i \sigma_2, \sigma_3\}$ operations
on his qubit to convert its state  to that of the unknown qubit a.

Let us now consider the case, where Alice makes two successive Bell 
measurements, i.e., von Neumann measurements using the Bell basis:
\begin{eqnarray}
\ket{\varphi^{\pm}} & =  & \frac{1}{\sqrt{2}}(\ket{00}\pm\ket{11}) \nonumber \\
\ket{\psi^{\pm}} & =  & \frac{1}{\sqrt{2}}(\ket{01}\pm\ket{10})
\end{eqnarray}

In the three-party scenario here, Alice will have qubits a and 1, Charlie
will have the qubits 2 and 3, while Bob would have the qubit 4.
In either of the two scenarios, the teleportation would be possible
but with different classical communication cost. It can be seen by 
rewriting equation (8) as,
\begin{eqnarray}
\ket{\psi}_a \ket{GHZ}_{1234} & = & \frac{1}{\sqrt{2}}(\alpha \ket{00}_{a1}\ket{00}_{23} \ket{0}_{4}+ 
    \alpha \ket{01}_{a1}\ket{11}_{23}\ket{1}_{4}+\nonumber \\
 &   &   \hspace{.3in}         \beta \ket{10}_{a1}\ket{00}_{23}\ket{0}_{4}+
 \beta \ket{11}_{a1}\ket{11}_{23}\ket{1}_{4} )                    \nonumber                \\ 
& = & \frac{1}{2 \sqrt{2}}[(\ket{\varphi^{+}}_{a1} \ket{\varphi^{+}}_{23} +
                \ket{\varphi^{-}}_{a1} \ket{\varphi^{-}}_{23})
                 (\alpha \ket{0}_{4}+ \beta \ket{1}_{4}) +  \nonumber \\
   & &   \hspace{.4in}   (\ket{\varphi^{+}}_{a1} \ket{\varphi^{-}}_{23} + \ket{\varphi^{-}}_{a1} \ket{\varphi^{+}}_{23})
                (\alpha  \ket{0}_{4}-\beta \ket{1})_{4} + \nonumber  \\
   & &    \hspace{.4in} (\ket{\psi^{+}}_{a1} \ket{\varphi^{+}}_{23} - \ket{\psi^{-}}_{a1} \ket{\varphi^{-}}_{23})
                (\alpha \ket{1}_{4}+\beta  \ket{0})_{4}+ \nonumber \\
   & &       \hspace{.4in} (\ket{\psi^{-}}_{a1} \ket{\varphi^{+}}_{23} - \ket{\psi^{+}}_{a1} \ket{\varphi^{-}}_{23})
                (\alpha \ket{1}_{4}-\beta \ket{0}_{4})].
\end{eqnarray}

    Here classical communication cost would be same as in the last
scenario  - two cbits if there are two parties and three cbits if there are
three parties.

    For this entangled resource, there is one more possibility. In this case 
Alice first makes a Bell measurement on the particles a and 1, followed by 
one-particles measurement on the particles 3 and 4. Or, there could be three 
or four parties. The distribution of the particles in the three-party scenario 
will be as above. In the four-party scenario, Alice would have particles a and 1,
Charlie would have particle 2 and Dennis would have particle 3, whereas
Bob would have particle 4. One can easily check that in these scenarios,
the teleportation is also possible because one can write $\ket{3GHZ_{1}^\pm}$ and
$\ket{3GHZ_{2}^\pm}$ in (12) in terms of the Bell states (15) and  the single-particle 
states $\ket{\pm}$ given in (14). The classical information cost would 
depend on the number of parties. For two parties, it would be 2 cbits; for
three parties, it would be 3 cbits; whereas for four parties, it would be
4 cbits.



\subsubsection{Teleportation using $\ket{\Omega}$ state}

     As noted earlier, this is one of the most interesting quadripartite entangled 
state and most powerful. As before, 
Alice wishes to teleport the unknown state $\ket{\psi}$ to Bob. However, the quantum 
resource available to her is the state $\ket{\Omega}$. This state is shared 
by particles 1, 2, 3, and 4. Alice has the particles 1, 2, and 3. Bob has the
particle 4. The combined state of the the five particles a, 1, 2, 3, and 4 will be

\begin{eqnarray}
\ket{\psi}_a \ket{\Omega}_{1234} & =  & ( \alpha \ket{0}_{a}+ \beta \ket{1}_{a})\otimes(\frac{1}{\sqrt{2}}(\ket{0}_{1}\ket{\varphi^+}_{23}
\ket{0}_{4}+\ket{1}_{1}\ket{\varphi^-}_{23}\ket{1}_{4})  \nonumber \\
& = & \frac{1}{\sqrt{2}}(\alpha \ket{00}_{a1} \ket{\varphi^+}_{23}\ket{0}_{4}+ \alpha \ket{01}_{a1} \ket{\varphi^-}_{23} \ket{1}_{4}
 + \beta \ket{10}_{a1} \ket{\varphi^+}_{23} \ket{0}_{4} +  \nonumber \\
 & & \beta \ket{11}_{a1} \ket{\varphi^-}_{23} \ket{1}_{4}).
\end{eqnarray}

 One can rewrite this combined state as,

\begin{eqnarray}
\ket{\psi}_a \ket{\Omega}_{1234} & =  & \ket{\Omega_1^+}_{a123} (\alpha \ket{0}_{a} + \beta \ket{1}_{a})
      +\ket{\Omega_1^-}_{a123} (\alpha\ket{0}_{a} -  \beta \ket{1}_{a})+ \nonumber \\
   &   &  \ket{\Omega_2^+}_{a123}(\alpha\ket{1}_{a} +  \beta \ket{0}_{a})
+\ket{\Omega_2^-}_{a123} (\alpha \ket{1}_{a} - \beta \ket{0}_{a}).
\end{eqnarray}
 
     Here the basis vectors are:
\begin{eqnarray}
\ket{\Omega_1^{\pm}} & = & \frac{1}{\sqrt{2}}\ket{00}\ket{\varphi^+}\pm\ket{11}\ket{\varphi^-}, \nonumber \\
\ket{\Omega_2^{\pm}} & = & \frac{1}{\sqrt{2}}\ket{01}\ket{\varphi^-}\pm\ket{10}\ket{\varphi^+}.
\end{eqnarray}

    After making the measurement in this basis, Alice can convey her results to 
Bob using two classical bits.  Bob then can apply appropriate unitary transformation 
to his qubit, as in the case of the GHZ-state, and complete the protocol. Interestingly,
there exist another set of four-particle basis vectors that Alice can use to make
measurement. This basis, GHZ-basis,  has in addition to the states (9) and (10),
following states
\begin{eqnarray}
\ket{4GHZ_{3}^{\pm}} & = & \frac{1}{\sqrt{2}}(\ket{0011}\pm\ket{1100}), \\
\ket{4GHZ_{4}^{\pm}} & = & \frac{1}{\sqrt{2}}(\ket{0100}\pm\ket{1011}).
\end{eqnarray}

    In this GHZ-basis, the combined state (17) can be written as,
\begin{eqnarray}
\ket{\psi}_a \ket{\Omega}_{1234} & = & \frac{1}{2 \sqrt{2}}[(\ket{4GHZ_1^+}_{a123} +\ket{4GHZ_3^-}_{a123})
         (\alpha \ket{0}_{4} - \beta \ket{1}_{4}) + \nonumber \\
  &  &   (\ket{4GHZ_1^-} _{a123}+ \ket{4GHZ_3^+}_{a123} )
      (\alpha \ket{0}_{4} + \beta \ket{1}_{4}) + \nonumber \\
    & & (-\ket{4GHZ_2^-}_{a123} + \ket{4GHZ_4^+}_{a123})
              (\alpha \ket{1}_{4} + \beta \ket{0}_{4}) +  \nonumber \\
 &   &  (-\ket{4GHZ_2^+}_{a123} + \ket{4GHZ_4^-}_{a123})(\alpha \ket{1}_{4} - \beta \ket{0}_{4})].
\end{eqnarray}

      By encoding the unitary operations in the two cbits, Alice can convey
the information to Bob who can complete the protocol.

    Let us now consider the next scenario where Alice makes a three-particle
von Neumann measurement followed by a one-particle measurement. As in
the case of $\ket{GHZ}$, there can be two situations. There can be two
parties or three parties. In this case we can rewrite (17) as,

\begin{eqnarray}
\ket{\psi}_a \ket{\Omega}_{1234} & =  & ( \alpha \ket{0}_{a}+ \beta \ket{1}_{a})\otimes(\frac{1}{\sqrt{2}}(\ket{0}_{1}\ket{\varphi^+}_{23}
\ket{0}_{4}+\ket{1}_{1}\ket{\varphi^-}_{23}\ket{1}_{4})  \nonumber \\
& = & \frac{1}{4}[(\ket{3GHZ_{1}^{+}}_{a12} \ket{-}_{3} + \ket{3GHZ_{1}^{-}}_{a12} \ket{+}_{3} +
            \ket{3GHZ_{3}^{+}}_{a12} \ket{+}_{3} - \ket{3GHZ_{3}^{-}}_{a12} \ket{-}_{3})  \nonumber \\
   &&       \;\qquad       (\alpha \ket{0}_{4}+ \beta \ket{1}_{4}) +  \nonumber \\
   & &    (\ket{3GHZ_{1}^{+}}_{a12} \ket{+}_{3} + \ket{3GHZ_{1}^{-}}_{a12} \ket{-}_{3}   -
        \ket{3GHZ_{3}^{+}}_{a12} \ket{-}_{3} + \ket{3GHZ_{3}^{-}}_{a12} \ket{+}_{3})  \nonumber \\
   &&       \;\qquad        (\alpha  \ket{0}_{4}-\beta \ket{1})_{4} + \nonumber  \\
   & &   (\ket{3GHZ_{4}^{+}}_{a12} \ket{+}_{3} + \ket{3GHZ_{4}^{-}}_{a12} \ket{-}_{3}  +
          \ket{3GHZ_{2}^{+}}_{a12} \ket{-}_{3} - \ket{3GHZ_{2}^{-}}_{a12} \ket{+}_{3})  \nonumber \\
    & &     \;\qquad        (\alpha \ket{1}_{4}+\beta  \ket{0})_{4}+ \nonumber \\
   & &    (\ket{3GHZ_{4}^{+}}_{a12} \ket{-}_{3} + \ket{3GHZ_{4}^{-}}_{a12} \ket{+}_{3} -
   \ket{3GHZ_{2}^{+}}_{a12} \ket{+}_{3} + \ket{3GHZ_{2}^{-}}_{a12} \ket{-}_{3})  \nonumber \\
        & &   \;\qquad      (\alpha \ket{1}_{4}-\beta \ket{0}_{4})].
\end{eqnarray}

     Here one can use three-particle GHZ-basis. It has in addition to the states given in  
(12) and (13), we have the following states,

\begin{eqnarray}
\ket{3GHZ_3^{\pm}} & = & \frac{1}{\sqrt{2}}(\ket{001}\pm\ket{110}), \nonumber \\
 \ket{3GHZ_4^{\pm}} & = & \frac{1}{\sqrt{2}}(\ket{010}\pm\ket{101}).
\end{eqnarray}

As in the case of $\ket{GHZ}$ state, in the two-party situation,  Alice needs to send
two cbits to Bob (e.g., encoding the four unitary operations). In the three-party
situation, combined classical information cost will be three cbits. As before,
this classical communication could take many forms. For example, Charlie can send
one cbit of information about his measurement to Alice. On the basis of her
results, Alice can send two cbits of information to Bob, encoding the unitary 
transformation. 
After receiving the classical communication, Bob can complete the 
protocol by applying a suitable unitary operation.


The strategy of making two successive Bell measurements also works if we
 make measurements on suitably chosen qubits. (This is because, this
state is not symmetric under the permutation of  qubits.) If Alice makes a
measurement on qubits `a2' and `13', then we can write the combined
state as:

\begin{eqnarray}
\ket{\psi}_a \ket{\Omega}_{1234} & = & \frac{1}{\sqrt{2}}(\alpha \ket{0}_{a}  +
 \beta \ket{1}_{a}) (\ket{0}_{1} \ket{\varphi^+}_{23} \ket{0}_{4}+  \ket{1}_{1} \ket{\varphi^-}_{23}\ket{1}_{4} )                    \nonumber                \\ 
& = & \frac{1}{4}[( \ket{\varphi^{+}}_{a2} \ket{\varphi^{+}}_{13} + \ket{\varphi^{-}}_{a} \ket{\varphi^{-}}_{13} + \ket{\psi^{+}}_{a2} \ket{\psi^{-}}_{13} + \ket{\psi^{-}}_{a2} \ket{\psi^{+}}_{13})
                 (\alpha \ket{0}_{4} - \beta \ket{1}_{4}) +  \nonumber \\
&  &  (\ket{\varphi^{+}}_{a2} \ket{\varphi^{-}}_{13} + \ket{\varphi^{-}}_{a} \ket{\varphi^{+}}_{13} + \ket{\psi^{+}}_{a2} \ket{\psi^{+}}_{13} + \ket{\psi^{-}}_{a2} \ket{\psi^{-}}_{13})
                 (\alpha \ket{0}_{4}+ \beta \ket{1}_{4}) +  \nonumber \\
&  &  (\ket{\varphi^{+}}_{a2} \ket{\psi^{+}}_{13} - \ket{\varphi^{-}}_{a} \ket{\psi^{-}}_{13} + \ket{\psi^{+}}_{a2} \ket{\varphi^{-}}_{13} - \ket{\psi^{-}}_{a2} \ket{\varphi^{+}}_{13})
                 (\alpha \ket{1}_{4}+ \beta \ket{0}_{4}) -  \nonumber \\
&  &  (\ket{\varphi^{+}}_{a2} \ket{\psi^{-}}_{13} - \ket{\varphi^{-}}_{a} \ket{\psi^{+}}_{13} - \ket{\psi^{+}}_{a2} \ket{\varphi^{+}}_{13} + \ket{\psi^{-}}_{a2} \ket{\varphi^{+}}_{13})
                 (\alpha \ket{1}_{4} - \beta \ket{0}_{4})].
\end{eqnarray}

Here the classical communication cost will
be two cbits for two-party situation and four cbits in the three-party
situation. 

Other situation is a Bell measurement followed by two one-particle measurements.
As before, one could have two-party, three-party and four-party situations. The
protocol would work as before with appropriate classical communication cost. 
This is because one can rewrite $\ket{3GHZ_{n}^{\pm}}$ ($n = 1 - 4$) in
(23) in terms of the single-particle states (14) and the Bell states (15).

\subsubsection{Teleportation using $\ket{W}$ state}

    Next we consider the W-state given in (2). This state does not allow the faithful 
teleportation of an unknown qubit state. However, 
a modified version of this state that also belongs to the W-state category 
under the SLOCC classification can work. This is shown below.  The distribution 
of the four qubits is as earlier. The combined state of the particle
a, 1, 2, 3, and 4 can be written as

\begin{eqnarray}
\ket{\psi}_a \ket{W}_{1234} & =  & ( \alpha \ket{0}_{a}+ \beta \ket{1}_{a})\otimes \frac{1}{2}(\ket{1000}_{1234}+\ket{0100}_{1234}+\ket{0010}_{1234}+\ket{0001}_{1234}  )\nonumber \\
& = & \frac{1}{2}(\alpha \ket{0100}_{a123}\ket{0}_{4}  + \alpha \ket{0010}_{a123}\ket{0}_{4} +
       \alpha \ket{0001}_{a123}\ket{0}_{4} + \alpha \ket{0000}_{a123}\ket{1}_{4} + \nonumber \\
  & &    \; \beta \ket{1100}_{a123}\ket{0}_{4}  + \beta \ket{1010}_{a123}\ket{0}_{4} +
       \beta \ket{1001}_{a123}\ket{0}_{4} + \beta \ket{1000}_{a123}\ket{1}_{4} ).
\end{eqnarray}

It does not seem possible to rewrite the above state 
to teleport the qubit faithfully. In the case of three-qubit W-state, 
it was shown\cite{ap} that instead of the W state,
one needs to consider the state $\ket{W_{n}}$, which is 
 $\frac{1}{\sqrt{2+2n}}(\ket{100}+\sqrt{n}\ket{010}+
\sqrt{n+1}\ket{001})$. This state  can be used for the perfect teleportation of 
an unknown qubit. Analogously, one could construct the state for the case of four qubits.
We can consider the state
\begin{equation}
\ket{W_{mn}}=\frac{1}{\sqrt{2m + 2n + 2}}(\ket{1000}+ m e^{i \rho}\ket{0100}+
n e^{i \eta}\ket{0010}+\sqrt{m + n + 1}e^{i \sigma}\ket{0001}).
\end{equation}
    
 Here $m$ and $n$ are real numbers. For simplicity, one could set the phases to unity
 and choose $m = n = 1$,
\begin{equation}
\ket{W_{11}}=\frac{1}{\sqrt{6}}(\ket{1000}+\ket{0100}+\ket{0010}+\sqrt{3}\ket{0001}.
\end{equation}
  With this quantum resource, the combined state of five particles would be

\begin{eqnarray}
\ket{\psi}_a \ket{W_{11}}_{1234} & =  & \frac{1}{\sqrt{6}}( \alpha \ket{0}_{a}+ \beta \ket{1}_{a})\otimes (\ket{1000}_{1234}+\ket{0100}_{1234}+\ket{0010}_{1234}+\sqrt{3} \ket{0001}_{1234}  )\nonumber \\
& = & \frac{1}{\sqrt{6}}(\alpha \ket{0100}_{a123}\ket{0}_{4}  + \alpha \ket{0010}_{a123}\ket{0}_{4} +
       \alpha \ket{0001}_{a123}\ket{0}_{4} +\sqrt{3}  \alpha \ket{0000}_{a123}\ket{1}_{4} + \nonumber \\
  & &    \; \beta \ket{1100}_{a123}\ket{0}_{4}  + \beta \ket{1010}_{a123}\ket{0}_{4} +
       \beta \ket{1001}_{a123}\ket{0}_{4} + \sqrt{3} \beta \ket{1000}_{a123}\ket{1}_{4} ) \nonumber \\
& = & \frac{1}{\sqrt{6}}[\ket{\eta^+}_{a123}(\alpha \ket{0}_{4}+\beta \ket{1}_{4})+\ket{\eta^-}_{a123}(\alpha \ket{0}_{4}-\beta\ket{1}_{4}) + \nonumber \\
 &  &     \;\;\ket{\zeta^+}_{a123}(\beta\ket{0}_{4}+\alpha \ket{1}_{4})+\ket{\zeta^-}_{a123}(\alpha \ket{0}_{4} - \beta\ket{1}_{4},
\end{eqnarray}

where,

\begin{eqnarray}
\ket{\eta^{\pm}} & = & \frac{1}{\sqrt{6}}(\ket{0100}+\ket{0010}+\ket{0001}\pm
\sqrt{3}\ket{1000}), \nonumber \\
\ket{\zeta^{\pm}} & = & \frac{1}{\sqrt{6}}(\ket{1100}+\ket{1010}+\ket{1001}\pm
\sqrt{3}\ket{0000}). 
\end{eqnarray}

  Now Alice can send the two classical bits of information to Bob to inform him
 about the results of his measurement, or the unitary operation that he 
 should apply. Bob then completes the protocol
by applying appropriate unitary transformation.

One can make a more general observation about constructing a suitable state.
This state can be 
$p\ket{1000}+q\ket{0100}+r\ket{0010}+s\ket{0001}$ where
$|p|^2+|q|^2+|r|^2=|s|^2$ which is suitable for teleportation of qubit.
For a more general case of N-qubits the suitable state would be
$a_1\ket{10...0}+a_2\ket{010...0}+...+a_N\ket{00....1}$ with
coefficients satisfying $|a_1|^2+|a_2|^2+...+|a_{N-1}|^2=|a_N|^2$. 

As earlier, there also exist scenarios of Alice making three-particle 
measurement followed by one-particle measurement; two successive 
Bell measurements; one Bell measurement followed by two one-particle 
measurements. In all of these scenarios, there could exist multiparty
situations. These scenarios may work out with suitable W-class
state. As before, more parties would mean more classical communication
cost. 

\subsubsection {Teleportaion using  $\ket{Q4}$ state}

 The state $\ket{Q4}$ can also be used to teleport an unknown state $\ket{\psi}$.
 Unlike the GHZ-state, here the state changes with
permutation of the particles. The states obtained on permutation would also
belong to the same SLOCC class. However, for different states, one would need
different distribution of particles for the measurement.
If the particles are distributes such that particles
a, 1, 2, and 3 are with Alice and 4 with Bob then this state cannot be used for
teleportation; but if the distribution is such that  Alice has particles a, 1, 3,  4,
 and Bob has particle 2, it would  lead to faithful teleportation. 
We can see how the protocol works as follows. The combined state
of the five particles can be written as

\begin{eqnarray}
\ket{\psi}_a\ket{Q4}_{1234}  &  = &
\frac{1}{2}(\alpha \ket{0}_{a}+ \beta \ket{1}_{a})(\ket{0000}_{1234} +\ket{0101}_{1234}+\ket{1000}_{1234}+\ket{1110}_{1234}   \nonumber  \\
  & = & \frac{1}{2}[\alpha \ket{0000}_{a134}\ket{0}_{2}+ \alpha \ket{0001}_{a134}\ket{1}_{2}+ \alpha \ket{0100}_{a134}\ket{0}_{2}
+ \alpha \ket{0011}_{a134}\ket{1}_{2}+ \nonumber  \\
 &  &   \beta  \ket{1000}_{a134}\ket{0}_{2}+\beta  \ket{1001}_{a134}\ket{1}_{2}+\beta  
\ket{1100}_{a134}\ket{0}_{2} + \beta \ket{1011}_{a134}\ket{1}_{2}].
\end{eqnarray}

Now Alice can use one of the following set of basis vectors to make four-particle von-Neumann measurements.
One set of basis vectors are

\begin{eqnarray}
\ket{\rho_1^{\pm}} & = & \frac{1}{2}[(\ket{0000}+\ket{0100})\pm(\ket{1001}+\ket{1011})] \\
\ket{\rho_2^{\pm}} & = & \frac{1}{2}[(\ket{0001}+\ket{0011})\pm(\ket{1001}+\ket{1100})],
\end{eqnarray}

while the other set is,

\begin{eqnarray}
\ket{\tau_1^{\pm}} & = & \frac{1}{\sqrt{2}}(\ket{0000}\pm\ket{1001}) \\
\ket{\tau_2^{\pm}} & = & \frac{1}{\sqrt{2}}(\ket{0001}\pm\ket{1000}) \\
\ket{\tau_3^{\pm}} & = & \frac{1}{\sqrt{2}}(\ket{0100}\pm\ket{1011}) \\
\ket{\tau_4^{\pm}} & = & \frac{1}{\sqrt{2}}(\ket{0011}\pm\ket{1100}).
\end{eqnarray}

Using the basis (32)-(33), one can rewrite (31) as

\begin{eqnarray}
\ket{\psi}_a \ket{Q4}_{1234} & = & \frac{1}{2}[\ket{\rho_1^+}_{a134} (\alpha \ket{0}_{2} + \beta \ket{1}_{2})
+\ket{\rho_1^-}_{a134} (\alpha \ket{0}_{2}-\beta \ket{1}_{2})+ \nonumber \\
  & &  \ket{\rho_2^+}_{a134} (\alpha \ket{1}_{2}+\beta \ket{0}_{2})
+\ket{\rho_2^- }_{a134} (\alpha \ket{1}_{2}-\beta \ket{0}_{2})],
\end{eqnarray}

while using the basis (34)-(37), we can rewrite (31) as,

\begin{eqnarray}
\ket{\psi}_a \ket{Q4}_{1234} & = & \frac{1}{2\sqrt{2}}[(\ket{\tau_1^+} + \ket{\tau_3^+})(\alpha \ket{0}+\beta \ket{1})
+(\ket{\tau_1^-}+\ket{\tau_3^-})(\alpha \ket{0}-\beta \ket{1}) + \nonumber \\
 & &  (\ket{\tau_2^+}+\ket{\tau_4^+})(\alpha \ket{1}+\beta \ket{0})
+(\ket{\tau_2^-}+\ket{\tau_4^-})(\alpha \ket{1}-\beta \ket{0})].
\end{eqnarray}

Irrespective of the basis set Alice uses, she needs to send only two 
classical bits of information to Bob. Then, Bob
can apply suitable unitary operator to convert the state of his qubit to that of
(7).

It is interesting to note  that if the particles are distributed such that Alice has
particles a, 1, 2, 3 and Bob has 4, then these exists a state in this SLOCC class,

\begin{equation}
\ket{Q4_{11}}=\frac{1}{\sqrt{6}}(\ket{0000}+\ket{1000}+\ket{1110}+\sqrt{3}\ket{0101},
\end{equation}
which can be used for the  teleportation if  the measurement is performed in the basis,
\begin{eqnarray}
\ket{\eta}^{\pm} & = & \frac{1}{\sqrt{6}}(\ket{0000}+\ket{0100}+\ket{0111}\pm
\sqrt{3}\ket{1010})\\
\ket{\zeta}^{\pm} & = & \frac{1}{\sqrt{6}}(\ket{1000}+\ket{1100}+\ket{1111}\pm
\sqrt{3}\ket{0010}).
\end{eqnarray}

 As earlier, there exist the scenarios where
 Alice chooses to make a three-particle measurement followed by
a one-particle measurement; or she makes two successive Bell measurements; or she
makes a Bell measurement followed by two one-particle measurements.
These scenarios could have multiparty situations. One needs to investigate further
whether these scenarios could be realized with this $\ket{Q4}$ state. One
can also explore other states of this class for their suitability for realizing
various scenarios.

\subsubsection{Teleportation using $\ket{Q5}$ state}

   This entangled state can also be used as a suitable quantum resource.
 The distribution of the four qubits is as before. The combined state of the particle
a, 1, 2, 3, and 4 can be written as:

\begin{eqnarray}
\ket{\psi}_a \ket{Q5}_{1234} & =  & ( \alpha \ket{0}_{a}+ \beta \ket{1}_{a})\otimes \frac{1}{2}(\ket{0000}_{1234}+\ket{1011}_{1234}+\ket{1101}_{1234}+\ket{1110}_{1234}  )\nonumber \\
& = & \frac{1}{2}(\alpha \ket{0000}_{a123}\ket{0}_{4}  + \alpha \ket{0101}_{a123}\ket{1}_{4} +
       \alpha \ket{0110}_{a123}\ket{1}_{4} + \alpha \ket{0111}_{a123}\ket{0}_{4} + \nonumber \\
  & &    \; \beta \ket{1000}_{a123}\ket{0}_{4}  + \beta \ket{1101}_{a123}\ket{1}_{4} +
       \beta \ket{1110}_{a123}\ket{1}_{4} + \beta \ket{1111}_{a123}\ket{0}_{4} ).
\end{eqnarray}

     It turns out that one can teleport the state $\ket{\psi}$, if Alice makes a four-particle
 von Neumann measurement using at least two different sets of basis vectors.
If Alice uses the following set of basis vectors:

\begin{eqnarray}
\ket{\varphi_1^{\pm}} & = & \frac{1}{2}[(\ket{0000}+\ket{0111})\pm(\ket{1101}+\ket{1110})]\\
\ket{\varphi_2^{\pm}} & = & \frac{1}{2}[(\ket{0101}+\ket{0110})\pm(\ket{1000}+\ket{1111})],
 \end{eqnarray}
then one can rewrite the combined state (43) as,

\begin{eqnarray}
\ket{\psi}_a \ket{Q5}_{1234} & = & \frac{1}{2}[\ket{\varphi_1^+}_{a123}(\alpha \ket{0}_{4}+\beta \ket{1}_{4})+\ket{\varphi_1^-}_{a123}(\alpha \ket{0}_{4}-\beta\ket{1}_{4}) + \nonumber \\
  &  & \ket{\varphi_2^+}_{a123}(\beta\ket{0}_{4}+\alpha \ket{1}_{4})+\ket{\varphi_2^-}_{a123}(\alpha \ket{0}_{4} - \beta\ket{1}_{4}],
\end{eqnarray}

or, one can use the basis vectors,

\begin{eqnarray}
\ket{\xi_1^{\pm}} & = & \frac{1}{\sqrt{2}}(\ket{0000}\pm\ket{1101}) \\
\ket{\xi_2^{\pm}} & = & \frac{1}{\sqrt{2}}(\ket{0111}\pm\ket{1110}) \\
\ket{\xi_3^{\pm}} & = & \frac{1}{\sqrt{2}}(\ket{0101}\pm\ket{1000}) \\
\ket{\xi_4^{\pm}} & = & \frac{1}{\sqrt{2}}(\ket{0110}\pm\ket{1111}),
\end{eqnarray}

then the combined state would be,

\begin{eqnarray}
\ket{\psi}_a \ket{Q5}_{1234} & = & \frac{1}{2\sqrt{2}}[(\ket{\xi_1^+}_{a123} +\ket{\xi_2^+}_{a123})
                   (\alpha \ket{0}_{4}+\beta \ket{1}_{4})    + (\ket{\xi_1^-}_{a123} +\ket{\xi_2^-}_{a123})
                   (\alpha \ket{0}_{4}-\beta \ket{1}_{4}) +  \nonumber \\
    & &               (\ket{\xi_3^+}_{a123} +\ket{\xi_4^+}_{a123})
                   (\alpha \ket{1}_{4}+\beta \ket{0}_{4})    + (\ket{\xi_3^-}_{a123} +\ket{\xi_4^-}_{a123})
                   (\alpha \ket{1}_{4}-\beta \ket{0}_{4})].
\end{eqnarray}

Whatever the basis set, Alice uses to make von Neumann measurements,
she needs to send two classical bit of information to Bob. She could encode
the operations that Bob should apply. On receiving the classical information,
Bob can complete the protocol.

Let us now consider the scenario, when Alice makes a three-particle followed by a
one-particle measurement. Unlike the GHZ-state, here the state changes with
permutation of the particles. The states obtained on permutation would also
belong to the same SLOCC class. However, for different states, one would need
different distribution of particles for the measurement. In this specific case,
Alice needs to make a measurement on the particles a, 2, and 3 followed by that on
particle 1.

For the three-particle measurement, Alice can use the basis,

\begin{eqnarray}
\ket{\omega_1^{\pm}} & = & \frac{1}{\sqrt{2}}(\ket{000}\pm\ket{101}) \\
\ket{\omega_2^{\pm}} & = & \frac{1}{\sqrt{2}}(\ket{001}\pm\ket{100})\\
\ket{\omega_3^{\pm}} & = & \frac{1}{\sqrt{2}}(\ket{010}\pm\ket{111})\\
\ket{\omega_4^{\pm}} & = & \frac{1}{\sqrt{2}}(\ket{011}\pm\ket{110}).
\end{eqnarray}

Then the combined state (43) can be written as
\begin{eqnarray}
\ket{\psi}_a \ket{Q5}_{1234} & = & \frac{1}{4} [(\ket{\omega_1^+}_{a23} \ket{+}_{1} + \ket{\omega_1^-}_{a23}\ket{-}_{1} +\ket{\omega_4^+}_{a23}
\ket{+}_{1} - \ket{\omega_4^+}_{a23}  \ket{-}_{1})(\alpha \ket{0}_{4}+ \beta \ket{1}_{4}) + \nonumber \\
    &  &    (\ket{\omega_1^+}_{a23} \ket{-}_{1} +
                                    \ket{\omega_1^-}_{a23} \ket{+}_{1} +\ket{\omega_4^-}_{a23}
               \ket{+}_{1} - \ket{\omega_4^-}_{a23} \ket{-}_{1})
                            (\alpha \ket{0}_{4} -  \beta \ket{1}_{4}) + \nonumber \\
      & &                       ( \ket{\omega_2^+}_{a23} \ket{+}_{1} -
                                    \ket{\omega_2^-}_{a23}\ket{-}_{1} +\ket{\omega_3^+}_{a23}
\ket{+}_{1} - \ket{\omega_3^+}_{a23}  \ket{-}_{1})(\alpha \ket{1}_{4}+ \beta \ket{0}_{4}) + \nonumber \\
    &  &    (\ket{\omega_2^-}_{a23} \ket{+}_{1} -
                                    \ket{\omega_2^+}_{a23} \ket{-}_{1} +\ket{\omega_3^-}_{a23}
\ket{+}_{1} - \ket{\omega_3^-}_{a23} \ket{-}_{1})](\alpha \ket{0}_{4}-  \beta \ket{1}_{4})].
\end{eqnarray}

In two-party situation in this scenario, using the basis (52)-(55) for the 
measurement and using two classical bits of information the protocol can 
be carried out. For the three-party situation, the three cbits of classical
communication would be required. For the other scenarios, which we 
considered for the $\ket{GHZ}$ and $\ket{\Omega}$ states, one needs
to explore further to find the feasibility of the use of this category
of states.

\subsection{Teleportation of two-qubit state}

    In this section, we consider the possibility of teleporting an unknown arbitrary two-qubit
state. As we shall see that it would be possible with only a few quadripartite 
entangled states. However, sometime one would be able to teleport subclasses
of the general state. 

Let us consider the scenario where Alice has an unknown two-qubit state
of the particles a and b,
\begin{equation}
\ket{\psi}_{ab} = \alpha \ket{00}_{ab}+\beta \ket{01}_{ab}+
                             \gamma \ket{10}_{ab}+ \delta \ket{11}_{ab}.
\end{equation}

Here $\alpha$,  $\beta$, $\gamma$ and $\delta$ are complex numbers.
She wishes to transmit this state to Bob using a quadripartite entangled state.
So Bob will have two of the four qubits and Alice will have the other two. As in
the teleportation protocol, Alice can make von Neumann measurement on her qubits and
communicate her results to Bob  using a classical channel. Bob then attempts to create
this state using unitary transformations on his two qubits. As before, Alice may 
have many bases to choose from. There can also exist many scenarios. In the
first scenario, Alice makes a four-particle measurement and communicates
classically with Bob. In the second scenario, Alice makes a three-particle
measurements, followed by one particle measurement and classical
communication. In this second scenario, one could have two-party and
three-party situations. In a third scenario, Alice may choose to make
two suitable Bell measurements, instead of a four-particle measurement.
  
If one considers  a product state of two Bell-states as a quantum resource,
or the state $\ket{\chi}$ of Ref \cite{yc}, then one can teleport an arbitrary 
unknown two-qubit state. We shall now consider various quantum entangled 
resources as in the last section. It is shown that the cluster state $\ket{\Omega}$ 
may also be 
used for an arbitrary two-qubit state, however for other considered quantum
resources, only a sub-class can be teleported.


\subsubsection{\bf Teleportation of two-qubit state using $\ket{GHZ}$ state}

The GHZ state is not a  suitable entangled quantum resource for the teleportation 
of an arbitrary unknown two-qubit state.  However, an entangled two-qubit state of 
the the following form  can be teleported with the GHZ state (1),

\begin{equation}
\ket{\psi_{1}}_{ab}  = \sigma_i\sigma_j(\alpha \ket{00}_{ab} + \beta\ket{11}_{ab}.
\end{equation}

Here $ \sigma_i$  and $\sigma_j$ are Pauli matrix operators, while $\alpha$ and $\beta$
are complex numbers. We can see this by rewriting the combined state of the six 
particles as:

\begin{eqnarray}
\ket{\psi_{1}}_{ab} \ket{GHZ}_{1234} & = & \sigma_i\sigma_j(\alpha \ket{0000}_{ab12} \ket{00}_{34}
           + \alpha \ket{0011}_{ab12} \ket{11}_{34}   +  \nonumber \\
    & &      \beta \ket{1100}_{ab12}\ket{00}_{34} + \beta \ket{1111}_{ab12} \ket{11}_{34}       \nonumber \\
   &  = & \ket{\pi_1^+}_{ab12} (\alpha\ket{00}_{34}+\beta \ket{11}_{34})+\ket{\pi_1^-}_{ab12}
             (\alpha\ket{00}_{34}-\beta \ket{11}_{34}) + \nonumber \\
   & & \ket{\pi_2^+}_{ab12}  (\alpha \ket{11}_{34}+\beta\ket{00}_{34})+\ket{\pi_2^-}_{ab12} 
           (\alpha \ket{11}_{34}-\beta \ket{00}_{34}), 
\end{eqnarray}

where $\ket{\pi_{1,2}^{\pm}} $ are the orthogonal states,
\begin{eqnarray}
\ket{\pi_1^{\pm}}& =& \sigma_i\sigma_j\ket{4GHZ_1^{\pm}}   \nonumber \\
\ket{\pi_2^{\pm}}& = & \sigma_i\sigma_j\ket{4GHZ_3^{\pm}}.
\end{eqnarray}

After Alice has made the measurement in this basis and conveyed the results
to Bob using two classical bits, he can carry out unitary operations 
$(\sigma_i\sigma_0)(\sigma_j\sigma_0),
(\sigma_i\sigma_0)(\sigma_j\sigma_3),(\sigma_i\sigma_1)(\sigma_j\sigma_1)$, or
$(\sigma_i\sigma_1)(\sigma_j\sigma_2)$ for the corresponding results 
of $\ket{\pi}_1^+,\ket{\pi}_1^-,\ket{\pi}_2^+$ and $\ket{\pi}_2^-$. 
This way Bob can obtain the state (59). 

Instead of making a four-particle measurement, Alice may choose to make a
three-particle on the particles a, b, and 1,  followed by a one-particle measurement
on the particle 2. This will also result in the successful teleportation, except that
now Alice will have to convey three classical bits of information to Bob.
There is also a possibility of Alice making two Bell measurements on the
particles a and 1,  followed by on the particles b and 2.  These situations need to
be explored further.

\subsubsection{Teleportation of two-qubit state using $\ket{\Omega}$ state}

It turns out that the cluster state $\ket{\Omega}$ can be a quite important
entangled quantum resource. An arbitrary two-qubit state can be teleported 
using this state. To see this we can rewrite the combined state of the systems
given in (3) and (57) as

\begin{eqnarray}
\ket{\psi}_{ab}\ket{\Omega}_{1234} & = & \frac{1}{2}
    [(\alpha \ket{0000}_{ab12}\ket{01}_{34}+\alpha \ket{0001}_{ab12}\ket{11}_{34}+\alpha \ket{0010}_{ab12}\ket{00}_{34}-\alpha \ket{0011}_{ab12}\ket{10}_{34})+ \nonumber \\
   & & (\beta \ket{0100}_{ab12}\ket{01}_{34}+\beta \ket{0101}_{ab12}\ket{11}_{34}+\beta \ket{0110}_{ab12}\ket{00}_{34}-\beta \ket{0111}_{ab12}\ket{10}_{34})+ \nonumber \\
  & &  \gamma \ket{1000}_{ab12}\ket{01}_{34}+\gamma \ket{1001}_{ab12}\ket{11}_{34}+\gamma \ket{1010}_{ab12}\ket{00}_{34}-\gamma \ket{1011}_{ab12}\ket{10}_{34})+ \nonumber \\
  & &  \delta \ket{1100}_{ab12}\ket{01}_{34}+\delta \ket{1101}_{ab12}\ket{11}_{34}+\delta \ket{1110}_{ab12}\ket{00}_{34}-\delta \ket{1111}_{ab12}\ket{10}_{34})].
\end{eqnarray}

Let us know consider the possibility of Alice making a four-particle measurement
and then conveying her results to Bob. Bob then makes suitable unitary transformation
on his two qubits. It turns out that following set of basis vectors can serve as a suitable
measurement basis,

\begin{eqnarray}
\ket{\Omega_1} & = & \frac{1}{2}(\ket{0000}+\ket{0110}+\ket{1001}-\ket{1111}) \\
\ket{\Omega_2} & = & \frac{1}{2}(\ket{0000}+\ket{0110}-\ket{1001}+\ket{1111}) \\
\ket{\Omega_3} & = & \frac{1}{2}(\ket{0000}-\ket{0110}+\ket{1001}+\ket{1111})\\
\ket{\Omega_4}  &  = & \frac{1}{2}(\ket{0000}-\ket{0110}-\ket{1001}-\ket{1111}) \\
\ket{\Omega_5} &  = & \frac{1}{2}(\ket{0001}+\ket{0111}+\ket{1000}-\ket{1110}) \\
\ket{\Omega_6}  & = & \frac{1}{2}(\ket{0001}+\ket{0111}-\ket{1000}+\ket{1110})\\
\ket{\Omega_7}  & =& \frac{1}{2}(\ket{0001}-\ket{0111}+\ket{1000}+\ket{1110}) \\
\ket{\Omega_8}  &= &\frac{1}{2}(\ket{0001}-\ket{0111}-\ket{1000}-\ket{1110})\\
\ket{\Omega_9}  &= &\frac{1}{2}(\ket{0010}+\ket{0100}+\ket{1011}-\ket{1101})\\
\ket{\Omega_{10}} & = &\frac{1}{2}(\ket{0010}+\ket{0100}-\ket{1011}+\ket{1101}) \\
\ket{\Omega_{11}} & = &\frac{1}{2}(\ket{0010}-\ket{0100}+\ket{1011}+\ket{1101}) \\
\ket{\Omega_{12}} & = &\frac{1}{2}(\ket{0010}-\ket{0100}-\ket{1011}-\ket{1101})\\
\ket{\Omega_{13}}  &= & \frac{1}{2}(\ket{0011}+\ket{0101}-\ket{1010}+\ket{1100})\\
\ket{\Omega_{14}} & = &\frac{1}{2}(\ket{0011}+\ket{0101}+\ket{1010}-\ket{1100}) \\
\ket{\Omega_{15}} & = &\frac{1}{2}(\ket{0011}+\ket{0101}-\ket{1010}+\ket{1100})\\ 
\ket{\Omega_{16}} & = &\frac{1}{2}(\ket{0011}-\ket{0101}-\ket{1010}-\ket{1100}).
\end{eqnarray}

We can then rewrite (61) as,
\begin{eqnarray}
\ket{\psi}_{ab}\ket{\Omega}_{1234} & = & \frac{1}{4}[
\ket{\Omega_1}_{ab12}(\alpha \ket{01}_{34}+\beta \ket{00}_{34}+\gamma \ket{11}_{34}+\delta \ket{10}_{34}) \nonumber \\
&&+\ket{\Omega_2}_{ab12}(\alpha \ket{01}_{34}+\beta \ket{00}_{34}-\gamma \ket{11}_{34}-\delta \ket{10}_{34}) \nonumber \\
&&+\ket{\Omega_3}_{ab12}(\alpha \ket{01}_{34}-\beta \ket{00}_{34}+\gamma \ket{11}_{34}-\delta \ket{10}_{34}) \nonumber \\
&&+\ket{\Omega_4}_{ab12}(\alpha \ket{01}_{34}-\beta \ket{00}_{34}-\gamma \ket{11}_{34}+\delta \ket{10}_{34})  \nonumber \\
&&+\ket{\Omega_5}_{ab12}(\alpha \ket{11}_{34}-\beta \ket{10}_{34}+\gamma \ket{01}_{34}-\delta \ket{10}_{34})\nonumber \\
 & & +\ket{\Omega_6}_{ab12}(\alpha \ket{11}_{34}-\beta \ket{10}_{34}-\gamma \ket{01}_{34}+\delta \ket{10}_{34})\nonumber \\
 & & +\ket{\Omega_7}_{ab12}(\alpha \ket{11}_{34}+\beta \ket{10}_{34}+\gamma \ket{01}_{34}+\delta \ket{10}_{34})\nonumber \\
 & & +\ket{\Omega_8}_{ab12}(\alpha \ket{11}_{34}+\beta \ket{10}_{34}-\gamma \ket{01}_{34}-\delta \ket{10}_{34})\nonumber \\
 & & +\ket{\Omega_9}_{ab12}(\alpha \ket{00}_{34}+\beta \ket{01}_{34}-\gamma \ket{10}_{34}-\delta \ket{11}_{34})\nonumber \\
 & & +\ket{\Omega_{10}}_{ab12}(\alpha \ket{00}_{34}+\beta \ket{01}_{34}+\gamma \ket{10}_{34}+\delta \ket{11}_{34})\nonumber \\
 & & +\ket{\Omega_{11}}_{ab12}(\alpha \ket{00}_{34}-\beta \ket{01}_{34}-\gamma \ket{10}_{34}+\delta \ket{11}_{34})\nonumber \\
 & & +\ket{\Omega_{12}}_{ab12}(\alpha \ket{00}_{34}-\beta \ket{01}_{34}+\gamma \ket{10}_{34}-\delta \ket{11}_{34})\nonumber \\
 & & -\ket{\Omega_{13}}_{ab12}(\alpha \ket{10}_{34}-\beta \ket{11}_{34}-\gamma \ket{10}_{34}+\delta \ket{11}_{34})\nonumber \\
 & & -\ket{\Omega_{14}}_{ab12}(\alpha \ket{10}_{34}-\beta \ket{11}_{34}+\gamma \ket{10}_{34}-\delta \ket{11}_{34})\nonumber \\
 & & -\ket{\Omega_{15}}_{ab12}(\alpha \ket{10}_{34}+\beta \ket{11}_{34}-\gamma \ket{10}_{34}-\delta \ket{11}_{34})\nonumber \\
 & & -\ket{\Omega_{16}}_{ab12}(\alpha \ket{10}_{34}+\beta \ket{11}_{34}+\gamma \ket{10}_{34}+\delta \ket{11}_{34})].
\end{eqnarray}

  Alice can now communicate four cbits of information to Bob. Bob can then recover the unknown state
 using appropriate unitary transformations. For example,  if after the measurement
the state of particles a,b,1, and 2 is 
$(\alpha \ket{01}+\beta \ket{00}+\gamma \ket{11}+\delta \ket{10})$, then Bob performs
$(\sigma_0\otimes\sigma_1)$ operation on his particles to recover the general state (57).

There exist another scenario where Alice makes separate Bell measurement on (a,2) and (b,1)
particles. To see what happens we can rewrite the state (61) as

\begin{eqnarray}
\ket{\psi}_{ab}\ket{\Omega}_{1234} & = & \frac{1}{4}[
\ket{\varphi^+}_{a2}\ket{\varphi^+}_{b1}(\alpha \ket{01}_{34}+\beta \ket{00}+\gamma \ket{11}-\delta \ket{10}) +  \nonumber \\
 & & \ket{\varphi^+}_{a2}\ket{\varphi^-}_{b1}(\alpha \ket{01}_{34}-\beta \ket{00}+\gamma \ket{11}-\delta \ket{10}) +  \nonumber \\
 & & \ket{\varphi^-}_{a2}\ket{\varphi^+}_{b1}(\alpha \ket{01}_{34}+\beta \ket{00}-\gamma \ket{11}-\delta \ket{10}) +  \nonumber \\
 & & \ket{\varphi^-}_{a2}\ket{\varphi^-}_{b1}(\alpha \ket{01}_{34}-\beta \ket{00}_{34}-\gamma \ket{11}_{34}+\delta \ket{10})_{34} +  \nonumber \\
 & & \ket{\psi^+}_{a2}\ket{\varphi^+}_{b1}(\alpha \ket{11}_{34}-\beta \ket{10}_{34}+\gamma \ket{01}_{34}+\delta \ket{10}_{34})+   \nonumber \\
 & & \ket{\psi^+}_{a2}\ket{\varphi^-}_{b1}(\alpha \ket{11}_{34}+\beta \ket{10}_{34}+\gamma \ket{01}_{34}-\delta \ket{10}_{34}) +   \nonumber \\
& & \ket{\psi^-}_{a2}\ket{\varphi^+}_{b1}(\alpha \ket{11}_{34}-\beta \ket{10}_{34}-\gamma \ket{01}_{34}-\delta \ket{10}_{34})+   \nonumber \\
& & \ket{\psi^-}_{a2}\ket{\varphi^-}_{b1}(\alpha \ket{11}_{34}+\beta \ket{10}_{34}-\gamma \ket{01}_{34}+\delta \ket{10}_{34})+   \nonumber \\
& & \ket{\varphi^+}_{a2}\ket{\psi^+}_{b1}(\alpha \ket{00}_{34}+\beta \ket{01}_{34}-\gamma \ket{10}_{34}+\delta \ket{11}_{34})+   \nonumber \\
& &\ket{\varphi^+}_{a2}\ket{\psi^-}_{b1}(\alpha \ket{00}_{34}-\beta \ket{01}_{34}-\gamma \ket{10}_{34}-\delta \ket{11}_{34})+   \nonumber \\
& &\ket{\varphi^-}_{a2}\ket{\psi^+_{b1}}(\alpha \ket{00}_{34}+\beta \ket{01}_{34}-\gamma \ket{10}_{34}+\delta \ket{11}_{34})+   \nonumber \\
& &\ket{\varphi^-}_{a2}\ket{\psi^-}_{b1}(\alpha \ket{00}_{34}-\beta \ket{01}_{34}+\gamma \ket{10}_{34}+\delta \ket{11}_{34})+   \nonumber \\
& &\ket{\psi^+}_{a2}\ket{\psi^+}_{b1}(-\alpha \ket{10}_{34}+\beta \ket{11}_{34}+\gamma \ket{00}_{34}+\delta \ket{01}_{34})+   \nonumber \\
& &\ket{\psi^+}_{a2}\ket{\psi^-}_{b1}(-\alpha \ket{10}_{34}-\beta \ket{11}_{34}+\gamma \ket{00}_{34}-\delta \ket{01}_{34})+   \nonumber \\
& &\ket{\psi^-}_{a2}\ket{\psi^+}_{b1}(-\alpha \ket{10}_{34}+\beta \ket{11}_{34}-\gamma \ket{00_}{34}-\delta \ket{01}_{34})+   \nonumber \\
& &\ket{\psi^-}_{a2}\ket{\psi^-}_{b1}(-\alpha \ket{10}_{34}-\beta \ket{11}_{34}-\gamma \ket{00}_{34}+\delta \ket{01}_{34}).
\end{eqnarray}

In this case,
unitary operations by Bob on his individual qubits does not allow reconstruction of
the original state.  However, a controlled phase shift operation on particles 3 and 4, 
with the particle 3 as the control bit and 4 as the target bit, followed by a unitary operation 
yield the exact state. For example, if the measurement results of the particle (a,2) and (b,1) are 
$\ket{\varphi^+}$ and $\ket{\varphi^+}$, respectively the state of the particle collapse into
the state $\alpha \ket{01}+\beta \ket{00}+\gamma \ket{11}-\delta \ket{10}$. 
This state can not be transformed
into $\ket{\psi}_{ab}$ by individual unitary transformations only. A controlled
 phase operation transforms above state into $\alpha \ket{01}+\beta \ket{00}-\gamma \ket{11}
-\delta \ket{10}$ which can be converted into $\ket{\psi}_{ab}$ by $(\sigma_3\otimes\sigma_1)$. 
In the above protocol, use of the Bell measurement constraints perfect teleportation by
nonlocal operation by Bob. 

\subsubsection{Teleportation of two qubit state using $\ket{W}$}

A general unknown two-qubit state cannot be teleported using $\ket{W}$-state as a quantum
resource. However, as in the case of the $\ket{GHZ}$-state as a quantum resource,
an unknown state like $\alpha \ket{00}+\beta \ket{11}$ may be teleported if Bob could make
entangled unitary transformations \cite{lq}. It may be possible to teleport a more general 
two-qubit state with a suitable state belonging to the W-state category under SLOCC. For
example,  probabilistic teleportation of the two-qubit state $\alpha \ket{00}+\beta \ket{01} + \beta \ket{10}$
is possible using the entangled resource as the prototype W-state. 



\subsubsection{Teleportation of two qubit state using $\ket{Q4}$}

While arbitrary two-qubit state cannot be teleported with the $\ket{Q4}$ state,
one may be able to teleport a subclass of states such as $\alpha\ket{00}+\beta \ket{01}+\gamma \ket{10}$ 
with suitable choice of basis and distribution of particles. Here Bob may need to use
 entangled unitary transformations.


\subsubsection{Teleportation of two qubit state using $\ket{Q5}$}

As before, the teleportation of an arbitrary unknown two-qubit state may not be possible.
However,  for a subclass of states, teleportation may be possible. For example, the state
 $\alpha \ket{00}+\beta \ket{10}+\gamma \ket{11}$, where the particles are
distributed such that the particles 2 and 3 are with Alice and 1 and 4 are with Bob
may be teleported  if Bob could apply  entangled unitary transformations.


\subsection {Teleportation of three-qubit state}

For the teleportation protocol, one last possibility that one may consider 
is an arbitrary and unknown three-qubit state $\ket{\psi}_{abc}$. In this
scenario, Alice has access to one particle
(1 or 2 or 3 or 4), while Bob has rest of the three particles. Alice performs
a joint measurement on four particles with suitable basis and transmits
her result to Bob who makes suitable unitary transformations on his three 
qubits.

It is clear that a quadripartite state would not be suitable for the teleportation
of a general unknown three-qubit state.  For the teleportation of such a state, one would
need an entangled state of six qubits. For example, three Bell states together
could be used for such a teleportation. Here Alice and Bob would share one qubit
each of the three Bell pair. This quantum resource does not have genuine six-particle
entanglement, but some other such states like cluster states may also work.

With quadripartite states, one would be able to teleport some subclasses of the
the general three-qubit state. We illustrate this by giving a few examples of
such subclasses.

\subsubsection{\bf Teleportation of three-qubit state using $\ket{GHZ}$}

With this state as a quantum resource, for the teleportation we consider the
following states:

\begin{equation}
 \ket{\psi_{2}}_{abc} = \sigma_i \sigma_j \sigma_k (\alpha \ket{000}_{abc}+\beta \ket{111}_{abc}),
\end{equation}
where  $i, j, k = 0, 1, 2, 3$ and $\sigma_i,\sigma_j$ and $\sigma_k$ act on particle 1, 2 and 3 respectively.

The joint state of this three-qubit state and GHZ state can be written as,

\begin{eqnarray}
\ket{\psi_{2}}_{abc} \ket{GHZ}_{1234} & = & \sigma_i\sigma_j\sigma_k(\alpha \ket{0000}_{abc1} \ket{000}_{234}+ \alpha \ket{0001}_{abc1} \ket{111}_{234}+ \nonumber \\
  &  &  \beta \ket{1110}_{abc1} \ket{000}_{234}+\beta \ket{1111}_{abc1}\ket{111}_{234}) \nonumber \\
   & =   &  \ket{\pi_3^+}_{abc1}(\alpha \ket{000}_{234}+\beta \ket{111}_{234}+\ket{\pi_3^-}_{abc1}(\alpha \ket{000}_{234}-\beta \ket{111})_{234}
+ \nonumber  \\
 & & \ket{\pi_4^+}_{abc1}(\alpha \ket{111}_{234}+\beta \ket{000}_{234})+\ket{\pi_4^-}_{abc1}(\alpha \ket{111}_{234}-\beta \ket{000}_{234})
\end{eqnarray}

Therefore if Alice performs a measurement in the basis,
\begin{eqnarray}
\ket{\pi_3^{\pm}} & = & \frac{1}{\sqrt{2}}\sigma_i\sigma_j\sigma_k
\ket{0000}\pm\ket{1111})  \\
   \ket{\pi_4^{\pm}} & = & \frac{1}{\sqrt{2}}\sigma_i\sigma_j\sigma_k
\ket{0001}\pm\ket{1110})
\end{eqnarray}
and sends two classical bits of information to Bob, then Bob can apply  unitary transformations
$(\sigma_i\sigma_0)(\sigma_j\sigma_0)(\sigma_k\sigma_0),
(\sigma_i\sigma_0)(\sigma_j\sigma_0)(\sigma_k\sigma_3),
(\sigma_i\sigma_0)(\sigma_j\sigma_0)(\sigma_k\sigma_1),
(\sigma_i\sigma_0)(\sigma_j\sigma_0)(\sigma_k\sigma_2)$
to retrieve the original three-qubit state.

\subsubsection{\bf Teleportation of three-qubit state using $\ket{\Omega}$}

  With this quantum resource, one teleport following subclass of two-qubit
state:

\begin{equation}
 \ket{\psi_{3}}_{abc} = \sigma_i \sigma_j  (\alpha \ket{\varphi^+}_{ab}\ket{1}_{c} +\beta \ket{\varphi^-}_{ab}\ket{0}_{c}),
\end{equation}
where  $i, j = 0, 1, 2, 3$ and $\sigma_i$ acts on the particles a or b, while $\sigma_j$  act on particle c.

We can rewrite this combined state as:
\begin{eqnarray}
 \ket{\psi_{3}}_{abc} (\ket{0}_{1} \ket{\varphi^+}_{23} \ket{0}_{4} +\ket{1}_{1} \ket{\varphi^-}_{23}\ket{1}_{4}) &  = &
\frac{1}{\sqrt{2}}\sigma_i\sigma_j[\ket{\varphi^+}_{ab}\ket{10}_{c1} \alpha \ket{\varphi^+}_{23}\ket{0}_{4}
+ \nonumber \\ 
 & & \ket{\varphi^+}_{ab}\ket{11}_{c1}\alpha \ket{\varphi^-}_{23}\ket{1}_{4}  
 +  \ket{\varphi^-}_{ab}\ket{00}_{c1}\beta \ket{\varphi^+}_{23} \ket{0}_{4} \nonumber \\
& &  + \:\ket{\varphi^-}_{ab}\ket{01}_{c1}\beta \ket{\varphi^-}_{23}\ket{1}_{4}] \nonumber \\
& = &  \ket{\Omega_3^+}_{abc1}(\alpha \ket{\varphi^+}_{23}\ket{0}_{4}+\beta \ket{\varphi^-}_{23}\ket{1}_{4}) + \nonumber \\
 &  & \ket{\Omega_3^-}_{abc1}(\alpha \ket{\varphi^+}_{23}\ket{0}_{4}-\beta \ket{\varphi^-}_{23}\ket{1}_{4}) + \nonumber \\
 & & \ket{\Omega_4^+}_{abc1}(\alpha \ket{\varphi^-}_{23}\ket{1}_{4}+\beta \ket{\varphi^+}_{23}\ket{0}_{4}) + \nonumber \\
 & &  \ket{\Omega_4^-}_{abc1}(\alpha \ket{\varphi^+}_{23}\ket{1}_{4} - \beta \ket{\varphi^+}_{23} \ket{0}_{4}),
\end{eqnarray}


If Alice makes the measurement in the basis
\begin{eqnarray}
\ket{\Omega_3^{\pm}} & = & \frac{1}{\sqrt{2}}\sigma_{i} \sigma_{j}(\ket{\varphi^+}\ket{00} \pm\ket{\varphi^-}\ket{11}) \nonumber \\
\ket{\Omega_4^{\pm}} & = &\frac{1}{\sqrt{2}}\sigma_{i} \sigma_{j}(\ket{\varphi^+}\ket{10}\pm\ket{\varphi^-}\ket{01}),
\end{eqnarray}

then as we see, the state (84) can be teleported, once Alice sends two classical bits of
information to Bob.



\subsubsection{\bf Teleportation of three-qubit state using $\ket{W}$}

A three particle GHZ-state can be teleported with four particle GHZ state.
Can the three-particle W-states be teleported with the four-particle W-state ?
Let us consider general three particle W state

\begin{equation}
\ket{\psi_4}  =  \alpha \ket{001}+\beta \ket{010}+\gamma \ket{100}+\delta \ket{000}
\end{equation}

We can rewrite the combined state as,
\begin{eqnarray}
\ket{\psi_4}_{abc}\ket{W}_{1234} &   \nonumber \\
   = & \alpha \ket{0010}_{abc1}\ket{001}_{234}+\alpha \ket{0010}_{abc1}\ket{010}_{234}+\alpha \ket{0010}_{abc1}\ket{100}_{234}+\alpha \ket{0011}_{abc1}\ket{000}_{234} +  \nonumber \\
  & \beta \ket{0100}_{abc1}\ket{001}_{234}+\beta \ket{0100}_{abc1}\ket{010}_{234}+\beta \ket{0100}_{abc1}\ket{100}_{234}+\beta \ket{0101}_{abc1}\ket{000}_{234} + \nonumber \\
 & \gamma \ket{1000}_{abc1}\ket{001}_{234}+\gamma \ket{1000}_{abc1}\ket{010}_{234}+\gamma \ket{1000}_{abc1}\ket{100}_{234}+\gamma \ket{1001}_{abc1}\ket{000}_{234} + \nonumber \\
 & \delta \ket{0000}_{abc1}\ket{001}_{234}+\delta \ket{0000}_{abc1}\ket{010}_{234}+\delta \ket{0000}_{abc1}\ket{100}_{234}+\delta \ket{0001}_{abc1}\ket{000}_{234}.
\end{eqnarray}

No orthogonal measurement would enable faithful teleportation of $ \ket{\psi_4}$ state. However,
if we consider the coefficients such as  $\alpha = \beta =\gamma=\delta = \frac{1}{2}$ and 
Alice performs the  measurement in the basis

\begin{eqnarray}
\ket{\Sigma_1^{\pm}} & = & \frac{1}{\sqrt{2}}(\ket{0010}\pm\ket{0011}), \nonumber \\
\ket{\Sigma_2^{\pm}} & = & \frac{1}{\sqrt{2}}(\ket{0100}\pm\ket{0101}), \nonumber \\
\ket{\Sigma_3^{\pm}} & = & \frac{1}{\sqrt{2}}(\ket{1000}\pm\ket{1001}), \nonumber \\
\ket{\Sigma_1^{\pm}} & = & \frac{1}{\sqrt{2}}(\ket{0000}\pm\ket{0001},
\end{eqnarray}
then Bob's three particle will in one of the states,
$\frac{1}{2}(\ket{001}+\ket{010}+\ket{100}+\ket{000})$ or 
$\frac{1}{2}(\ket{001}+\ket{010}+\ket{100}-\ket{000})$.
Here the second state can not be transformed into the original state by local unitary
operation. However, a joint operation
 $\ket{001}\bra{001}+\ket{010}\bra{010}+\ket{100}\bra{100}
-\ket{000}\bra{000}$ could enable Bob to reproduce the original state. Here
classical communication cost is one cbit.

\section{Superdense coding}
     
       Superdense coding protocol has played an important role in the development
  of the field of Quantum Information. It was shown
  in \cite{bw} that using an entangled state as a resource, the classical
   capacity of a quantum channel can be enhanced. Suppose there are two parties -
  Alice and Bob. If Alice sends a qubit to Bob, then Bob can extract only
  one classical bit of information. However, if they share a Bell state, then using
  the protocol, Alice can transmit two classical bits by sending one qubit. Below,
   we are studying this protocol in the context of quadripartite entangled states.
   We consider two scenarios. In the first scenario, single-receiver dense coding,
   there are only two parties - Alice and Bob. This is the conventional scenario.
   In the second scenario,  multi-receiver dense coding, there are more than two 
   parties. There is one sender and multiple receivers. If these  receivers 
   could make global  operations, then this scenario would reduce to the single-receiver 
   scenario. Therefore these receivers are restricted to make only local  operations
   and communicate classically with one another.

\subsection{Single-receiver superdense coding}

In this section, we discuss the superdense coding capacity of various entangled 
states where a sender, Alice, sends either one, two, or three qubits to a receiver, 
Bob. We use the notations $DC_1, DC_2$, or $DC_3$ respectively for the scenarios
when one, two, or three qubits are sent. In $DC_1$, two parties Alice and 
Bob share one of the above four-qubit entangled states such that Alice
possesses one particle, say 1, and Bob possesses three particles 2, 3, 
and 4. Similarly, in $DC_2$, Alice possesses two particles 1 and 2, while Bob 
possesses 3 and 4. In $DC_3$, Alice has three particles 1, 2, and 3, whereas Bob 
has particle 4. For all such distributions, they follow the standard superdense 
coding protocol to transmit a classical message from Alice to Bob. In this
protocol, Alice applies unitary operations $\textit{I}, \sigma_1, i\sigma_2, \sigma_3$
with equal probabilities on her particles and sends them to Bob. Bob 
performs a joint measurement on all the four particles to retrieve the 
original message. It is to be noted that as some of the channels are 
asymmetric with respect to permutation of particles, the distribution 
may affect the superdense coding capacity of the states. But we 
will always consider the particle distribution for the superdense coding capacity 
to be maximal. Since orthogonal states can be perfectly distinguished
with some suitable measurement basis, in principle the deterministic dense coding 
capacity mainly depends on how many orthogonal states are obtained 
by unitary encoding on Alice's side. Therefore, here we seek to find out the 
number, N, of orthogonal states obtained (amount of information 
is $log_{2}N$) by unitary operations on the particles by the sender.

\subsubsection{\bf The $\ket{GHZ}$ state}

Let us consider the case, when the shared resource state is the $\ket{GHZ}$ state.
When Alice applies unitary operation on particle 1, it produces following states,
\begin{eqnarray}
I\otimes I\otimes I\otimes I\ket{GHZ} \rightarrow\frac{1}{\sqrt{2}}(\ket{0000}+\ket{1111})\nonumber \\
\sigma_1\otimes I\otimes I\otimes I\ket{GHZ} \rightarrow\frac{1}{\sqrt{2}}(\ket{1000}+\ket{0111})\nonumber \\
i\sigma_2\otimes I\otimes I\otimes I\ket{GHZ} \rightarrow\frac{1}{\sqrt{2}}(\ket{1000}-\ket{0111})\nonumber \\
\sigma_3 \otimes I\otimes I\otimes I\ket{GHZ} \rightarrow\frac{1}{\sqrt{2}}(\ket{0000}-\ket{1111}).
\end{eqnarray}

These states are mutually orthogonal and can be unambiguously 
distinguished. After receiving the qubit from Alice, Bob performs a joint 
four-particle von Neumann measurement in the four-particle basis 
$\{\ket{4GHZ_{1}^{\pm}}, \ket{4GHZ_{2}^{\pm}}\}$ to distinguish 
these states. In this way he acquires two bits of classical information
by receiving only one qubit. Similarly in $DC_2$ scenario,
Alice applies unitary operations on two particles. It gives rise to
sixteen states out which eight are orthogonal. In this process Bob 
can access only three cbits - not four cbits - by receiving two qubit 
by measuring in a proper $\{ \ket{4GHZ_n^{\pm}} \}$ basis. In $DC_3$
scenario, Alice applies unitary operations on her three qubits yielding
 sixteen orthogonal states. It leads to perfect transmission of four classical
bits of information. In general using N particle GHZ state one may be able
to send N bits of classical information by sending N-1 particles.   

\subsubsection{\bf The $\ket{W}$ state}

In the case of parties sharing a W-state, in the $DC_1$ scenario,
unitary operations on qubit 1 give the following states, 
\begin{eqnarray}
I\otimes I\otimes I\otimes I \ket{W} \rightarrow\frac{1}{2}(\ket{0001}+\ket{0010}+\ket{0100}+\ket{1000})\nonumber \\
\sigma_1 \otimes I\otimes I\otimes I \ket{W} \rightarrow\frac{1}{2}(\ket{1001}+\ket{1010}+\ket{1100}+\ket{0000})\nonumber\\
i\sigma_2 \otimes I\otimes I\otimes I \ket{W} \rightarrow\frac{1}{2}(-\ket{1001}-\ket{1010}-\ket{1100}+\ket{1000})\nonumber\\
\sigma_3 \otimes I\otimes I\otimes I \ket{W} \rightarrow\frac{1}{2}(\ket{0001}+\ket{0010}+\ket{0100}-\ket{1000}).
\end{eqnarray}
These states are not orthogonal and unambiguously discrimination is not 
possible. Therefore, Bob's measurement would not allow him to perfectly 
distinguish these states and transmission of two cbits by sending one 
qubit can not be achieved with unit probability. However, four-qubit 
W-state $\ket{W_{mn}}$ can be used to send two cbits by transmitting
one qubit. 

In the $DC_2$ scenario, the $\ket{W}$ state can actually be useful.
When Alice applies unitary transformation on two particles,  
it gives rise to following orthogonal states,
\begin{eqnarray}
I\otimes I\otimes I\otimes I\ket{W}\rightarrow\frac{1}{2}(\ket{0001}+\ket{0010}+\ket{0100}+\ket{1000}),\nonumber\\
\sigma_3\otimes \sigma_3\otimes I\otimes I\ket{W}\rightarrow\frac{1}{2}(\ket{0001}+\ket{0010}-\ket{0100}-\ket{1000}),\nonumber\\
\sigma_1\otimes I\otimes I\otimes I\ket{W}\rightarrow\frac{1}{2}(\ket{1001}+\ket{1010}+\ket{1100}+\ket{0000}),\nonumber\\
i\sigma_2\otimes \sigma_3\otimes I\otimes I\ket{W}\rightarrow\frac{1}{2}(-\ket{1001}-\ket{1010}+\ket{1100}+\ket{0000}),\nonumber\\
\sigma_3\otimes \sigma_1\otimes I\otimes I\ket{W}\rightarrow\frac{1}{2}(\ket{0101}+\ket{0110}+\ket{0000}-\ket{1100}),\nonumber\\
I\otimes i\sigma_2\otimes I\otimes I \ket{W} \rightarrow\frac{1}{2}(-\ket{0101}-\ket{0110}+\ket{0000}-\ket{1100}),\nonumber\\
\sigma_1\otimes i\sigma_2\otimes I\otimes I\ket{W}\rightarrow\frac{1}{2}(-\ket{1101}-\ket{1110}+\ket{1000}-\ket{0100}),\nonumber\\
i\sigma_2\otimes \sigma_1\otimes I\otimes I\ket{W}\rightarrow\frac{1}{2}(-\ket{1101}-\ket{1110}-\ket{1000}+\ket{0100}).
\end{eqnarray}
As these states are mutually orthogonal. Bob will be able to discriminate 
all these states with unit probability by von Neumann measurement 
and recover three cbits of information. The dense coding capacity for 
$DC_3$ scenario is limited to three cbits irrespective of conventional 
$\ket{W}$ state or $\ket{W_{mn}}$ as a quantum channel. It is 
interesting to note that a three particle W state $\frac{1}{\sqrt{3}}(\ket{001}+\ket{010}+\ket{100})$ is unsuitable for teleportation of 
a qubit and also one can not send more than one cbit of information 
using the superdense coding protocol.  But a four-particle W state of 
the above form though not suitable for a qubit teleportation, 
it can be used for superdense coding and one can transmit
three cbits by sending two qubits. The peculiarity of this state is that 
more than one cbit or three cbits of information cannot be transmitted
by sending one or three qubits respectively, but three cbits of information 
could be communicated with the transmission of two qubits. 

\subsubsection{The $\ket{\Omega}$ state}

This state is the best quantum resource from the point of view
of superdense coding. The classical information capacity in the
$DC_1$ scenario is two cbits with the proper distribution of the
qubits. However, in scenario $DC_2$, Alice can transmit four
cbits by transmitting two qubits. 
This makes the $\ket{\Omega}$
state as the best quantum resource of all the considered states.
In this scenario, when Alice applies unitary transformations on 
particles 1 and 2, following sixteen orthogonal states are obtained:
 
\begin{eqnarray}
I\otimes I\otimes I\otimes I\ket{\Omega}\rightarrow\frac{1}{2}(\ket{0000}+\ket{0110}+\ket{1001}-\ket{1111})\nonumber\\
I\otimes \sigma_3\otimes I\otimes I\ket{\Omega}\rightarrow\frac{1}{2}(\ket{0000}-\ket{0110}+\ket{1001}+\ket{1111})\nonumber\\ 
\sigma_3\otimes I\otimes I\otimes I\ket{\Omega}\rightarrow\frac{1}{2}(\ket{0000}+\ket{0110}-\ket{1001}+\ket{1111})\nonumber\\
\sigma_3\otimes \sigma_3\otimes I\otimes I\ket{\Omega}\rightarrow\frac{1}{2}(\ket{0000}-\ket{0110}-\ket{1001}-\ket{1111})\nonumber\\ 
I\otimes \sigma_1\otimes I\otimes I\ket{\Omega}\rightarrow\frac{1}{2}(\ket{0100}+\ket{0010}+\ket{1101}-\ket{1011})\nonumber\\
I\otimes \sigma_2\otimes I\otimes I\ket{\Omega}\rightarrow\frac{1}{2}(-\ket{0100}+\ket{0010}-\ket{1101}-\ket{1011})\nonumber\\
\sigma_3\otimes \sigma_1\otimes I\otimes I\ket{\Omega}\rightarrow\frac{1}{2}(\ket{0100}+\ket{0010}-\ket{1101}+\ket{1011})\nonumber\\
\sigma_3\otimes \sigma_2\otimes I\otimes I\ket{\Omega}\rightarrow\frac{1}{2}(-\ket{0100}+\ket{0010}+\ket{1101}+\ket{1011})\nonumber\\
\sigma_1\otimes I\otimes I\otimes I\ket{\Omega}\rightarrow\frac{1}{2}(\ket{1000}+\ket{1110}+\ket{0001}-\ket{0111})\nonumber\\
\sigma_1\otimes \sigma_3\otimes I\otimes I\ket{\Omega}\rightarrow\frac{1}{2}(\ket{1000}-\ket{1110}+\ket{0001}+\ket{0111})\nonumber\\
\sigma_2\otimes I\otimes I\otimes I\ket{\Omega}\rightarrow\frac{1}{2}(-\ket{1000}-\ket{1110}+\ket{0001}-\ket{0111})\nonumber\\
\sigma_2\otimes \sigma_3\otimes I\otimes I\ket{\Omega}\rightarrow\frac{1}{2}(-\ket{1000}+\ket{1110}+\ket{0001}+\ket{0111})\nonumber\\
\sigma_1\otimes \sigma_1\otimes I\otimes I\ket{\Omega}\rightarrow\frac{1}{2}(\ket{1100}+\ket{1010}+\ket{0101}-\ket{0011})\nonumber\\
\sigma_1\otimes \sigma_2\otimes I\otimes I\ket{\Omega}\rightarrow\frac{1}{2}(-\ket{1100}+\ket{1010}-\ket{0101}-\ket{0011})\nonumber\\
\sigma_2\otimes \sigma_1\otimes I\otimes I\ket{\Omega}\rightarrow\frac{1}{2}(-\ket{1100}-\ket{1010}+\ket{0101}-\ket{0011})\nonumber\\
\sigma_2\otimes \sigma_2\otimes I\otimes I\ket{\Omega}\rightarrow\frac{1}{2}(\ket{1100}-\ket{1010}-\ket{0101}-\ket{0011})
\end{eqnarray}

Therefore, clearly the classical capacity in this scenario is four cbits which
is the maximum possible. Thus, the $\ket{\Omega}$ state exhibits same 
information capacity as the tensor product of two Bell states.
In $DC_3$ scenario,  the classical information transmission 
is limited to four cbits only. With Quadripartite states, more than this
capacity is not possible as the Hilbert space of four qubits is sixteen
dimensional.

\subsubsection{The $\ket{Q4}$ state}

   This state is not symmetric under the permutation of particles. Therefore,
the success of the protocol depends on the distribution of the particles
between the parties. For example, let Alice has the qubit 1, whereas Bob
has the rest. Then on applying unitary operations on her qubit, the 
$\ket{Q4}$ can become,

\begin{eqnarray}
I\otimes I\otimes I\otimes I\ket{Q4}\rightarrow\frac{1}{2}(\ket{0000}+\ket{0101}+\ket{1000}+\ket{1110})\nonumber\\
\sigma_1\otimes I\otimes I\otimes I\ket{Q4}\rightarrow\frac{1}{2}(\ket{1000}+\ket{1101}+\ket{0000}+\ket{0110})\nonumber\\
\sigma_2\otimes I\otimes I\otimes I\ket{Q4}\rightarrow\frac{1}{2}(-\ket{1000}-\ket{1101}+\ket{0000}+\ket{0110})\nonumber\\
\sigma_3\otimes I\otimes I\otimes I\ket{Q4}\rightarrow\frac{1}{2}(\ket{0000}+\ket{0101}-\ket{1000}-\ket{1110}).
\end{eqnarray}

These states are not orthogonal to each other, so the protocol does
not succeed. However, if the particles are distributed such that Alice 
has the particle 2 and Bob has the particles 
1, 3 and 4, then unitary operation on the particle 2 yield four 
orthogonal states,
\begin{eqnarray}
I\otimes I\otimes I\otimes I\ket{Q4}\rightarrow\frac{1}{2}(\ket{0000}+\ket{0101}+\ket{1000}+\ket{1110})\nonumber\\
I\otimes \sigma_1\otimes I\otimes I\ket{Q4}\rightarrow\frac{1}{2}(\ket{0100}+\ket{0001}+\ket{1100}+\ket{1010})\nonumber\\
I\otimes \sigma_2\otimes I\otimes I\ket{Q4}\rightarrow\frac{1}{2}(-\ket{0100}+\ket{0001}-\ket{1100}+\ket{1010})\nonumber\\
I\otimes \sigma_3\otimes I\otimes I\ket{Q4}\rightarrow\frac{1}{2}(\ket{0000}+\ket{0101}-\ket{1000}-\ket{1110}).
\end{eqnarray}

    Therefore with this distribution of the particles, the protocol succeeds.
In the scenario $DC_{2}$, Alice applies unitary operations on the 
qubits 1 and 2. It gives rise to eight orthogonal states,
\begin{eqnarray}
I\otimes I\otimes I\otimes I\ket{Q4}\rightarrow\frac{1}{2}(\ket{0000}+\ket{0101}+\ket{1000}+\ket{1110})\nonumber\\
I\otimes \sigma_3\otimes I\otimes I\ket{Q4}\rightarrow\frac{1}{2}(\ket{0000}-\ket{0101}+\ket{1000}-\ket{1110})\nonumber\\
\sigma_3\otimes I\otimes I\otimes I\ket{Q4}\rightarrow\frac{1}{2}(\ket{0000}+\ket{0101}-\ket{1000}-\ket{1110})\\
\sigma_3\otimes \sigma_3\otimes I\otimes I\ket{Q4}\rightarrow\frac{1}{2}(\ket{0000}-\ket{0101}-\ket{1000}+\ket{1110})\nonumber\\
\sigma_1\otimes \sigma_1\otimes I\otimes I\ket{Q4}\rightarrow\frac{1}{2}(\ket{1100}+\ket{1001}+\ket{0100}+\ket{0010})\nonumber\\
\sigma_2\otimes \sigma_1\otimes I\otimes I\ket{Q4}\rightarrow\frac{1}{2}(-\ket{1100}-\ket{1001}+\ket{0100}+\ket{0010})\nonumber\\
\sigma_1\otimes \sigma_2\otimes I\otimes I\ket{Q4}\rightarrow\frac{1}{2}(-\ket{1100}+\ket{1001}-\ket{0100}+\ket{0010})\nonumber\\
\sigma_2\otimes \sigma_2\otimes I\otimes I\ket{Q4}\rightarrow\frac{1}{2}(\ket{1100}-\ket{1001}-\ket{0100}+\ket{0010}).
\end{eqnarray}

Therefore the dense coding capacity in $DC_2$ scenario is three cbits. 
The $DC_3$ scenario, the classical capacity of the  $\ket{Q4}$ state
is only three cbits, independent of particle distribution and choice of coefficients.

\subsubsection{The $\ket{Q5}$ state}
   
   As in the case of the $\ket{Q4}$ state, the distribution of the particles
is important. It turns out that the protocol can be implemented in the
$DC_{1}$ scenario. Alice can transmit two cbits by sending one qubit 
 when particles are distributed such that Alice has the particle 2 and 
Bob has the particles 1, 3 and 4. In this scenario, when Alice applies
unitary operations following orthogonal states are formed:

\begin{eqnarray}
I\otimes I\otimes I\otimes I \ket{Q5}\rightarrow\frac{1}{2}(\ket{0000}+\ket{1011}+\ket{1101}+\ket{1110})\nonumber\\
I\otimes \sigma_1 \otimes I\otimes I\ket{Q5}\rightarrow\frac{1}{2}(\ket{0100}+\ket{1111}+\ket{1001}+\ket{1010})\nonumber\\
I\otimes i\sigma_2 \otimes I\otimes I\ket{Q5}\rightarrow\frac{1}{2}(-\ket{0100}-\ket{1111}+\ket{1001}+\ket{1010})\nonumber\\
I\otimes \sigma_3 \otimes I\otimes I\ket{Q5}\rightarrow\frac{1}{2}(\ket{0000}+\ket{1011}-\ket{1101}-\ket{1110}).
\end{eqnarray}

In the $DC_2$ scenario, when Alice applies unitary transformations
on the qubits 1 and 2,  eight orthogonal states are obtained:

\begin{eqnarray}
I\otimes I\otimes I\otimes I\ket{Q5}\rightarrow\frac{1}{2}(\ket{0000}+\ket{1011}+\ket{1101}+\ket{1110})\nonumber\\
\sigma_1\otimes I\otimes I\otimes I\ket{Q5}\rightarrow\frac{1}{2}(\ket{1000}+\ket{0011}+\ket{0101}+\ket{0110})\nonumber\\
I\otimes \sigma_1\otimes I\otimes I\ket{Q5}\rightarrow\frac{1}{2}(\ket{0100}+\ket{1111}+\ket{1001}+\ket{1010})\nonumber\\
\sigma_1\otimes \sigma_1\otimes I\otimes I\ket{Q5}\rightarrow\frac{1}{2}(\ket{1100}+\ket{0111}+\ket{0001}+\ket{0010})\nonumber\\
I\otimes \sigma_2\otimes I\otimes I\ket{Q5}\rightarrow\frac{1}{2}(-\ket{0100}-\ket{1111}+\ket{1001}+\ket{1010})\nonumber\\
\sigma_1\otimes \sigma_2\otimes I\otimes I\ket{Q5}\rightarrow\frac{1}{2}(-\ket{1100}-\ket{0111}+\ket{0001}+\ket{0010})\nonumber\\
I\otimes \sigma_3\otimes I\otimes I\ket{Q5}\rightarrow\frac{1}{2}(\ket{0000}+\ket{1011}-\ket{1101}-\ket{1110})\nonumber\\
\sigma_1\otimes \sigma_3\otimes I\otimes I\ket{Q5}\rightarrow\frac{1}{2}(\ket{1000}+\ket{0011}-\ket{0101}-\ket{0110})
\end{eqnarray}

 Therefore in this scenario, Alice can transmit three cbits by sending
two qubits. One can also verify that in the $DC_3$ scenario, 
four cbits can be sent by Alice to Bob by the transmission of
three qubits. 

\subsection{Multi-receiver superdense coding}

   As noted earlier, in this protocol, there is one sender, but more than
one receiver. Alice wishes to transmit classical information to one of the
receivers with the assistance of  the other receiver. Therefore,
this protocol could also be called assisted superdense coding.
Consider the situation where Alice  shares an entangled state with the
receivers $B_1$ and $B_2$. If $B_1$ and $B_2$ combine 
together and make a global measurement then the situation 
is similar to that discussed above for a single-receiver. However in
the multi-receiver scenario, $B_1$ and $B_2$ perform measurements locally 
instead of global measurements and are allowed to communicate through 
a classical channel. In this subsection, we examine the 
usefulness of the considered quadripartite states for this protocol.

In the superdense coding, Alice can convert the shared entangled state
to a set of orthogonal  states  by applying suitable unitary operations
on her particles. In this scenario, the task of  $B_1$ and $B_2$  is to 
distinguish these orthogonal states by local operations and classical 
communications (LOCC). The subject of distinguishing the orthogonal 
states using LOCC has been discussed in the literature \cite{wshv,wh,cl}.
We would use these results.

Walgate et al \cite{wshv} have shown that any {\em two} orthogonal
multipartite states can be distinguished by LOCC. Walgate and Hardy \cite{wh}
generalized this result to a system of a qubit and a qudit. They showed
that if Alice has the qubit and she goes first, then a set of $\ell$ orthogonal
states $\ket{\psi^{i}}$ is LOCC distinguishable iff there is a basis
$\{\ket{0}_{A},\ket{1}_{A}\}$  for Alice to make measurement such that
in this basis
\begin{equation}
\ket{\psi^{i}}=\ket{0}_A\ket{\eta_0^i}_{B}+\ket{1}_A\ket{\eta_1^i}_B,
\end{equation}
where $\braket{\eta_0^i}{\eta_0^j}= \braket{\eta_1^i}{\eta_1^j}=0$ if $i \neq j$.
These indices take the values from $1$ to $\ell$. Chen and Li \cite{cl} generalized 
this result to the case of more general systems and found a condition for the
LOCC distinguishability of a set of orthogonal states. In particular, they showed
that  $\ell$ orthogonal
states $\{\ket{\Psi_i}\}$ is perfectly distinguishable by LOCC if there exists a set of
product vectors such that each state $\ket{\Psi_i}$ is superposition of some of these
product vectors as follows
\begin{equation}
\ket{\Psi_i}=\ket{\Phi_i^1}_A\ket{\xi_i^1}_B+...+\ket{\Phi_i^{m^i}}_A\ket{\xi_i^{m^i}}_B,
\end{equation}
and each product vector $\ket{\Phi_i^{k^i}}_A\ket{\xi_i^{k^i}}_B (1 \le k^i \le m^i)$
belongs to only a state $\ket{\Psi_i}$, i.e.,
\begin{eqnarray}
\bra{\Phi_i^{k^i}}\braket{\xi_i^{k^i}}{\Psi_j} & = & 0 \:   \forall \: i \ne j, \nonumber \\
\bra{\Phi_i^{k^i}}\braket{\xi_i^{k^i}}{\Psi_i} & \ne  & 0.
\end{eqnarray}

In other words if a set of multipartite possible states is LOCC distinguishable,
each possible state can be written as linear combination of product vectors such that
each product vector of a possible state is orthogonal to the other possible states.
From above theorem one can infer that in $2\otimes 2\otimes 2\otimes 2$ systems, for
bipartite splitting the number of orthogonal states that can be perfectly distinguished
by LOCC is bounded by the number of product states in the linear combination. For sixteen
orthogonal states to be LOCC distinguishable these have to be all product states.
Similarly at most four and eight orthogonal states can be distinguished perfectly
by LOCC if the set of states are linear combination of four and two product states
respectively.

\subsubsection{\bf The $\ket{GHZ}$ state}

Now consider the case when the parties share a four particle GHZ-state.
Here Alice wishes to convey classical information to $B_2$ (Bob-2) with the assistance of
$B_1$ (Bob-1). There are a number of ways to distribute qubits among Alice, $B_1$ and $B_2$.
Alice could have one or two qubits. If Alice has only one qubit, then she can convert
the shared quadripartite state into a set of four orthogonal states. As is the DC$_1$
scenario, in that case Alice would be able to transmit only two cbits to $B_2$, with
the assistance of $B_1$. Here we consider the case when Alice
possess particles 1 and 2 and perform unitary operations on these with equal probabilities.
She sends the particle 1 to $B_1$ and particle 2 to $B_2$. Then $B_1$ and $B_2$ share
the eight orthogonal states,
\begin{eqnarray}
 \ket{4GHZ_{1}^{\pm}}_{1234} =\frac{1}{\sqrt{2}}(\ket{00}_{13}\ket{00}_{24}\pm\ket{11}_{13}\ket{11}_{2
4}),\nonumber\\
 \ket{4GHZ_{2}^{\pm}}_{1234} =\frac{1}{\sqrt{2}}(\ket{01}_{13}\ket{11}_{24}\pm\ket{10}_{13}\ket{00}_{2
4}), \nonumber\\
 \ket{4GHZ_{3}^{\pm}}_{1234} =\frac{1}{\sqrt{2}}(\ket{01}_{13}\ket{01}_{24}\pm\ket{10}_{13}\ket{10}_{2
4}),\nonumber\\
 \ket{4GHZ_{4}^{\pm}}_{1234} =\frac{1}{\sqrt{2}} (\ket{11}_{13}\ket{01}_{24} \pm \ket{00}_{13}\ket{10}_{24}).
\end{eqnarray}

These states can be written in product decomposition using Bell basis,
\begin{eqnarray}
\ket{4GHZ_{1}^+}_{1234} & = & \frac{1}{\sqrt{2}}( \ket{\phi^+}_{13}\ket{\phi^+}_{24}+\ket{\phi^-}_{13}\ket{\phi^-}_{24}),\nonumber\\
\ket{4GHZ_{1}^-}_{1234} & = &\frac{1}{\sqrt{2}}  (\ket{\phi^+}_{13}\ket{\phi^-}_{24}+\ket{\phi^-}_{13}\ket{\phi^+}_{24}),\nonumber\\
\ket{4GHZ_{2}^+}_{1234} & = &\frac{1}{\sqrt{2}} (-\ket{\psi^+}_{13}\ket{\phi^+}_{24}+\ket{\psi^-}_{13}\ket{\phi^-}_{24}),\nonumber\\
\ket{4GHZ_{2}^-}_{1234} & = & \frac{1}{\sqrt{2}} (\ket{\psi^+}_{13}\ket{\phi^-}_{24}+\ket{\psi^-}_{13}\ket{\phi^+}_{24}),\nonumber\\
\ket{4GHZ_{3}^+}_{1234} & = &\frac{1}{\sqrt{2}}(\ket{\psi^+}_{13}\ket{\psi^+}_{24}+\ket{\psi^-}_{13}\ket{\psi^-}_{24}),\nonumber\\
\ket{4GHZ_{3}^-}_{1234} & = &\frac{1}{\sqrt{2}}(\ket{\psi^+}_{13}\ket{\psi^-}_{24}+\ket{\psi^-}_{13}\ket{\psi^+}_{24}),\nonumber\\
\ket{4GHZ_{4}^+}_{1234} & = &\frac{1}{\sqrt{2}}(\ket{\phi^+}_{13}\ket{\psi^+}_{24}-\ket{\phi^-}_{13}\ket{\psi^-}_{24}),\nonumber\\
\ket{4GHZ_{4}^-}_{1234} & = &\frac{1}{\sqrt{2}}(-\ket{\phi^+}_{13}\ket{\psi^-}_{24}-\ket{\phi^-}_{13}\ket{\psi^+}_{24}).
\end{eqnarray}

Here after receiving the qubit from the Alice, $B_1$ makes a measurement in the Bell
basis and conveys his results to $B_2$. $B_2$ then also makes a measurement on his
two qubits in the Bell basis. Depending on his results, he can distinguish all
the above eight orthogonal states and thus decipher the three cbits of the
information.

Let us now consider the another situation, where Alice sends her two qubits to
$B_2$ and none to $B_1$. After Alice applies unitary transformation, the eight 
orthogonal states would be as in (103). For the sake of convenience, let
us assume that Alice has the particles 1 and 2; $B_1$ has particle 3 and $B_2$ has
the particle 4. After Alice sends her two qubits to $B_2$, $B_2$ would have the
particles 1, 2, and 3.
The suitable decomposition of the states (103) can be written as

\begin{eqnarray}
\ket{4GHZ_{1}^+}_{1234}& = & \frac{1}{\sqrt{2}}( \ket{3GHZ_{1}^{+}}_{123}\ket{+}_{4}+\ket{3GHZ_{1}^-}_{123}\ket{-}_{4}),\nonumber\\
\ket{4GHZ_{1}^-}_{1234}& = & \frac{1}{\sqrt{2}}( \ket{3GHZ_{1}^{+}}_{123}\ket{-}_{4}+\ket{3GHZ_{1}^-}_{123}\ket{+}_{4}),\nonumber\\
\ket{4GHZ_{2}^+}_{1234}& = & \frac{1}{\sqrt{2}}( \ket{3GHZ_{2}^{+}}_{123}\ket{+}_{4}-\ket{3GHZ_{2}^-}_{123}\ket{-}_{4}),\nonumber\\
\ket{4GHZ_{2}^-}_{1234}& = & \frac{1}{\sqrt{2}}(- \ket{3GHZ_{2}^{+}}_{123}\ket{-}_{4}+\ket{3GHZ_{2}^-}_{123}\ket{+}_{4}),\nonumber\\
\ket{4GHZ_{3}^+}_{1234}& = & \frac{1}{\sqrt{2}}(  \ket{3GHZ_{3}^{+}}_{123}\ket{+}_{4}-\ket{3GHZ_{3}^-}_{123}\ket{-}_{4}),\nonumber\\
\ket{4GHZ_{3}^-}_{1234}& = & \frac{1}{\sqrt{2}}( -\ket{3GHZ_{3}^{+}}_{123}\ket{-}_{4}+\ket{3GHZ_{3}^-}_{123}\ket{+}_{4}),\nonumber\\
\ket{4GHZ_{4}^+}_{1234}& = & \frac{1}{\sqrt{2}}(  \ket{3GHZ_{4}^{+}}_{123}\ket{+}_{4}+\ket{3GHZ_{4}^-}_{123}\ket{-}_{4}),\nonumber\\
\ket{4GHZ_{4}^-}_{1234}& = & \frac{1}{\sqrt{2}}(  \ket{3GHZ_{4}^{+}}_{123}\ket{-}_{4}+\ket{3GHZ_{4}^-}_{123}\ket{+}_{4}).
\end{eqnarray}

After making a one-particle von Neumann measurement in the $\{ \ket{\pm} \}$ basis, $B_1$
communicates his results to $B_2$ using one cbit. $B_2$ can now make three-particle von Neumann
measurement using the $\{ \ket{3GHZ_{1}^{\pm}}, 
 \ket{3GHZ_{2}^{\pm}},  \ket{3GHZ_{2}^{\pm}},  \ket{3GHZ_{2}^{\pm}} \}$ basis and
decipher the state. In this way, Alice can communicate three cbits to $B_2$ with the assistance
of $B_1$. We notice that in this second scenario, $B_1$ has to use fewer cbits to communicate
with the $B_2$. This is the advantage of the second scenario over the first.

\subsubsection{\bf The $\ket{\Omega}$ state}

Let us now consider another quantum resource, the  $\ket{\Omega}$ state. 
As in the case of GHZ-state, if Alice has the particle 1 only, she can convert
the entangled resource to at most four orthogonal states,  and thus can send
two cbits to $B_2$ with the assistance of $B_1$. Let us again consider the
situation where Alice has two qubits, whereas $B_1$ and $B_2$ have one each.
By applying unitary transformations on qubits 1 and 2, Alice can transform the state into
sixteen orthogonal states. However, unlike in the GHZ -state, $B_1$ and $B_2$ can
distinguish only four orthogonal state using LOCC. This is because, the $\ket{\Omega}$
state is a superposition of four terms and the Hilbert space of four qubits is sixteen
dimensional. The states that can be distinguished are                                                                                                            
\begin{eqnarray}
\ket{\Omega}_{1}& = &\frac{1}{2}(\ket{00}_{13}\ket{00}_{24}+\ket{01}_{13}\ket{10}_{24}      +\ket{10}_{13}\ket{01}_{24}-\ket{11}_{13}\ket{11}_{24}),\nonumber\\
\ket{\Omega}_{5}& = &\frac{1}{2}(\ket{00}_{13}\ket{10}_{24}+\ket{01}_{13}\ket{00}_{24}
+\ket{10}_{13}\ket{11}_{24}-\ket{11}_{13}\ket{01}_{24}),\nonumber\\                         
 \ket{\Omega}_{9}& = &\frac{1}{2}(\ket{10}_{13}\ket{00}_{24}+\ket{11}_{13}\ket{10}_{24}      +\ket{00}_{13}\ket{01}_{24}-\ket{01}_{13}\ket{11}_{24}),nonumber\\                          
\ket{\Omega}_{13}& = &\frac{1}{2}(\ket{10}_{13}\ket{10}_{24}+\ket{11}_{13}\ket{00}_{24}     +\ket{00}_{13}\ket{11}_{24}-\ket{01}_{13}\ket{01}_{24}).
\end{eqnarray}

One can see that the measurement in basis $ \{\ket{00},\ket{01},\ket{10},\ket{11}\} $
by $B_1$ and $B_2$ would enable them to distinguish the four states. 
So using the $\ket{\Omega}$ state two cbits of information can be communicated in 
multi-receiver scenario. In the scenario where Alice sends her both qubits to $B_2$, the
result would be the same, i.e., Alice would be able to communicate only two cbits to $B_2$.           

\subsubsection{\bf The $\ket{W}$ state}

   This state is like the $\ket{\Omega}$ state in the sense that it is also a superposition
 of four terms. Therefore conclusions for various scenarios would be the same. In particular,
 we consider the case when Alice applies  unitary transformation on the qubits 1 and 2 of 
the shared W-state which give rise to eight orthogonal states given in (92). Alice then
sends one qubit each to $B_1$ and $B_2$.
The  four states which can be LOCC distinguishable are
\begin{eqnarray}
\ket{W_1}_{1234} & = &\frac{1}{2}(\ket{00}_{13}(\ket{01}+\ket{10})_{24}+(\ket{01}+\ket{10})_{13}
\ket{00}_{24},\nonumber\\
\ket{W_2}_{1234}& = &\frac{1}{2}(\ket{00}_{13}(\ket{01}-\ket{10})_{24}+(\ket{01}-\ket{10})_{13}
\ket{00}_{24},\nonumber\\
\ket{W_3}_{1234}& = &\frac{1}{2}(\ket{10}_{13}(\ket{01}+\ket{10})_{24}+(\ket{11}+\ket{00})_{13}
\ket{00}_{24},\nonumber\\
\ket{W_4}_{1234}& = &\frac{1}{2}(-\ket{10}_{13}(\ket{01}-\ket{10})_{24}+(-\ket{11}+\ket{00})_{13}
\ket{00}_{24}.
\end{eqnarray}

Writing these states in the Bell basis, we obtain
\begin{eqnarray}
\ket{W_1}_{1234}& = &(\ket{\phi^+}_{13} +\ket{\phi^-}_{13})\ket{\psi^+}_{24}+\ket{\psi^+}_{13}(\ket{\phi^+}_{24}+\ket{\phi^-}_{24}),\nonumber\\
\ket{W_2}_{1234}& = &(\ket{\phi^+}_{13}+\ket{\phi^-}_{13})\ket{\psi^-}_{24}+\ket{\psi^-}_{13}(\ket{\phi^+}_{24}+\ket{\phi^-}_{24}),\nonumber\\
\ket{W_3}_{1234}& = &(\ket{\psi^+}_{13}-\ket{\psi^-}_{13})\ket{\psi^+}_{24}+\ket{\phi^+}_{13}(\ket{\phi^+}_{24}+\ket{\phi^-}_{24}),\nonumber\\
\ket{W_4}_{1234}& = &(\ket{\psi^-}_{13}-\ket{\psi^+}_{13})\ket{\psi^-}_{24}+\ket{\phi^-}_{13}(\ket{\phi^+}_{24}+\ket{\phi^-}_{24}).
\end{eqnarray}

These states satisfy the criteria to be LOCC distinguishable. $B_1$ would make a measurement
in the Bell basis on his qubits and communicates the result to $B_2$. $B_2$ makes his own Bell measurement
and thus obtains two cbits of the information.

\subsubsection{\bf The $\ket{Q4}$ state}
    As in the above two cases, Alice can transmit at most two cbits to B2 using the
  multi-receiver protocol. When Alice has two qubits 1 and 2, then on applying the
unitary transformations, she can convert the $\ket{Q4}$ state into eight orthogonal 
states, given in (96). However, it appears that there does not exist an straightforward
way to distinguish even four of these orthogonal states by LOCC, because one cannot
easily put these states in the (101) form.

\subsubsection{\bf The $\ket{Q5}$ state}
     The situation about this state is like the $\ket{W}$ state. Alice can convert the
shared  $\ket{Q5}$ state into eight orthogonal states, given in (99). However, only
four states are LOCC distinguishable by  $B_1$ and $B_2$. One such set is,

\begin{eqnarray}
\ket{Q5_1}=\frac{1}{2}(\ket{00}_{13}\ket{00}_{24}+\ket{11}_{13}\ket{01}_{24}
+\ket{10}_{13}\ket{11}_{24}+\ket{11}_{13}\ket{10}_{24}),\nonumber\\
\ket{Q5_2}=\frac{1}{2}(\ket{10}_{13}\ket{00}_{24}+\ket{01}_{13}\ket{01}_{24}
+\ket{00}_{13}\ket{11}_{24}+\ket{01}_{24}\ket{10}_{24}),\nonumber\\
\ket{Q5_3}=\frac{1}{2}(\ket{00}_{13}\ket{10}_{24}+\ket{11}_{13}\ket{11}_{24}
+\ket{10}_{13}\ket{01}_{24}+\ket{11}_{13}\ket{00}_{24}),\nonumber\\
\ket{Q5_4}=\frac{1}{2}(\ket{10}_{13}\ket{10}_{24}+\ket{01}_{13}\ket{11}_{24}
+\ket{00}_{13}\ket{01}_{24}+\ket{01}_{13}\ket{00}_{24}).
\end{eqnarray}

$B_1$  makes von Neumann measurement in the basis $ \{\ket{00},\ket{01},\ket{10},\ket{11}\}$
and communicates his results to $B_2$.  $B_2$  also makes the von Neumann measurement
in the same basis and can distinguish the four states, thus obtaining the two cbits from Alice.

\section{Conclusion}
     In  this paper, we have considered a number of different genuine quadripartite
  entangled  states as quantum resources for the teleportation and the superdense
  coding. For the teleportation protocol, we have examined the possibility of transmitting
 one-qubit, two-qubit and three-qubit unknown quantum states. Apart from the conventional
  scenario, we have also considered  the multi-party scenarios and alternately
  the situations where Alice chooses to make a series of von Neumann measurements instead 
  of one von Neumann measurement. For the superdense coding, we have considered 
  the conventional single-receiver scenario as well as multi-receiver scenarios.

    We find that the cluster state   $\ket{\Omega}$ can be a very useful quantum resource.
  It can be used to teleport an arbitrary two-qubit unknown state. Using this state one can also 
  transmit four classical bits by sending two qubits.  In most of the other scenarios,
  this state is at least as good a resource as any other; often it is a better resource.
  Only in multi-receiver scenario this state is less successful than the $\ket{GHZ}$ state.
  This is because the LOCC distinguishability criteria requires that the resource should
  have minimal number of terms. This may even indicate that  the $\ket{\Omega}$ state
  has ``stronger'' entanglement.
   The  $\ket{GHZ}$ state is the next useful resource. One can use this state 
   and $\ket{\Omega}$  to transmit one-qubit state in all possible ways.
  However the state $\ket{GHZ}$ is not as useful as the $\ket{\Omega}$ state in the
   superdense coding and transmitting multiple-qubit states. Other quantum states, 
   $\ket{W}$,  $\ket{Q_{4}}$, and $\ket{Q_{5}}$ can also be useful resources in
  a number of scenarios. Their full utility needs to be investigated further.
   An interesting thread to explore will the use of entangled unitary transformations in the 
  implementation of various  quantum communication protocols. Another interesting
  avenue would be the use of higher-dimensional entangled states.

\end{document}